# Memlumor: a luminescent memory device for photonic neuromorphic computing

Alexandr Marunchenko[1*], Jitendra Kumar[1], Alexander Kiligaridis[1], Shraddha M. Rao[1], Dmitry Tatarinov[2], Ivan Matchenya[2], Elizaveta Sapozhnikova[2], Ran Ji[3,4], Oscar Telschow[3,4], Julius Brunner[3,4], Anatoly Pushkarev[2], Yana Vaynzof[3,4], Ivan G. Scheblykin[1*]

*Corresponding author(s). E-mail(s): shprotista@gmail.com; ivan.scheblykin@chemphys.lu.se;

[1] Chemical Physics and NanoLund, Lund University, P.O. Box 124, 22100 Lund, Sweden

[2] School of Physics and Engineering, ITMO University, 49 Kronverksky, St. Petersburg 197101, Russian Federation

[3] Chair for Emerging Electronic Technologies, Technical University of Dresden, Nöthnitzer Str. 61, 01187 Dresden, Germany

[4] Leibniz-Institute for Solid State and Materials Research Dresden, Helmholtzstraße 20, 01069 Dresden, Germany

## Abstract

Neuromorphic computing promises to transform the current paradigm of traditional computing towards Non-Von Neumann dynamic energy-efficient problem solving. Thus, dynamic memory devices capable of simultaneously performing nonlinear operations (volatile) similar to neurons and also storing information (non-volatile) alike brain synapses are in the great demand. To satisfy these demands, a neuromorphic platform has to possess intrinsic complexity reflected in the built-in diversity of its physical operation mechanisms. Herein, we propose and demonstrate the novel concept of a memlumor - an all-optical device combining memory and luminophore, and being mathematically a full equivalence of the electrically-driven memristor. By utilizing metal halide perovskites as a memlumor material platform, we demonstrate the synergetic coexistence of both volatile and non-volatile memory effects within a broad timescale from ns to days. We elucidate the origin of such complex response to be related to the phenomena of photodoping and photochemistry activated by a tunable light input and explore several possible realizations of memlumor computing. Leveraging on the existence of a history-dependent photoluminescent quantum yield in various material platforms, the memlumor device concept will trigger multiple new research directions in both material science and optoelectronics. We anticipate that the memlumor, as a new optical dynamic computing element, will add a new dimension to existing optical technologies enabling their transition into application in photonic neuromorphic computing.

Memristors are fundamental electrical components that make it possible combine information storage with computation. Originally proposed theoretically by Leon Chua in 1971,[1] memristors were first identified experimentally in the seminal work of Strukov and coworkers in 2008.[2] Memristors share the functionalities of synapses and neurons in a brain which makes them uniquely suited to serve as elementary building blocks for next generation bio-inspired neuromorphic computers.[3–5] These are projected to become superior in solving specific classes of computational problems[6] by replacing classical computers with a von Neumann architecture.

The concept of memristor has been extended onto a wider class of devices and systems. As an example, the photonics community has adopted the memristive paradigm for the conception of electro-optical and all-optically-driven memories.[7–12] These efforts are motivated by the advantages and new opportunities for the devices operating in presence of light, such as massive frequency-multiplexing and spatial parallelism supported by generous nonlinear dynamics present in optical systems.[13–16] Importantly, optical-based memristors can operate at the single-photon level,[17] opening the route toward room-temperature photonic quantum memristors.[18] Despite that most of the photonic memristors rely on hybrid electro-optical signal input[12,19] thus inheriting some of the disadvantages of the classical electrically-driven memristors, some reports demonstrated neuron-like features in all-optical memory devices.[20–23]

Still, the nascent concept of high-order neuromorphic computing, inspired by human brain, requires a photonics system complexity that combines both volatile and non-volatile memory across a broad scale of time and energy [6,7,24] – the complexity which has been has been rarely reported.[6,25] For such neuromorphic computing devices, the next generation architecture of neural networks - spiking neural networks, would enable superior energy efficiency as compared to that of traditional neural network architectures. Hence, to realize the full potential of this perspective computing paradigm, a hardware with an abundance of dynamic processes is required. Importantly, this hardware should combine most of the advantages provided by the optical-computing paradigm to offer energy-efficient computing capabilities.[18]

Any such novel system should share the similarity between the function of brain operating elements and classical electrically-driven memristors. In both cases, a set of time-dependent inputs (Fig. 1a) (external stimuli) is processed through a dynamic system that accounts for its history (Fig. 1b), resulting in the generation of a certain output (Fig.1c). In the case of a brain, the function is dual: to store the information (time-dependent inputs) and generate an output based on what was learned from previous interactions with the environment. To realize this, the brain consists of neurons and synapses (connections between neurons) as elementary building blocks (Fig. 1d) coupled into complex 3D networks.[26] The behavior of an electrically-driven (voltage-driven) memristor, as a direct analogy for a brain neuron operation, (Fig 1e) is formally defined by Chua as:[27]

$$J(t) = G(\vec{X}, V, t) V(t) \qquad (1)$$

$$\frac{d\vec{X}}{dt} = f(\vec{X}, V), \qquad (2)$$

where *J(t)* and *V(t)* are current (output) and voltage (input) connected via conductance $G(\vec{X},V,t)$ where $G(\vec{X},0,t) \neq \infty$. The vector $\vec{X}$ represents the internal state-variables that change upon the application of bias. In memristors, these variables reflect dynamic changes related to temperature, filament conductance, phase transitions and others.[28–32]

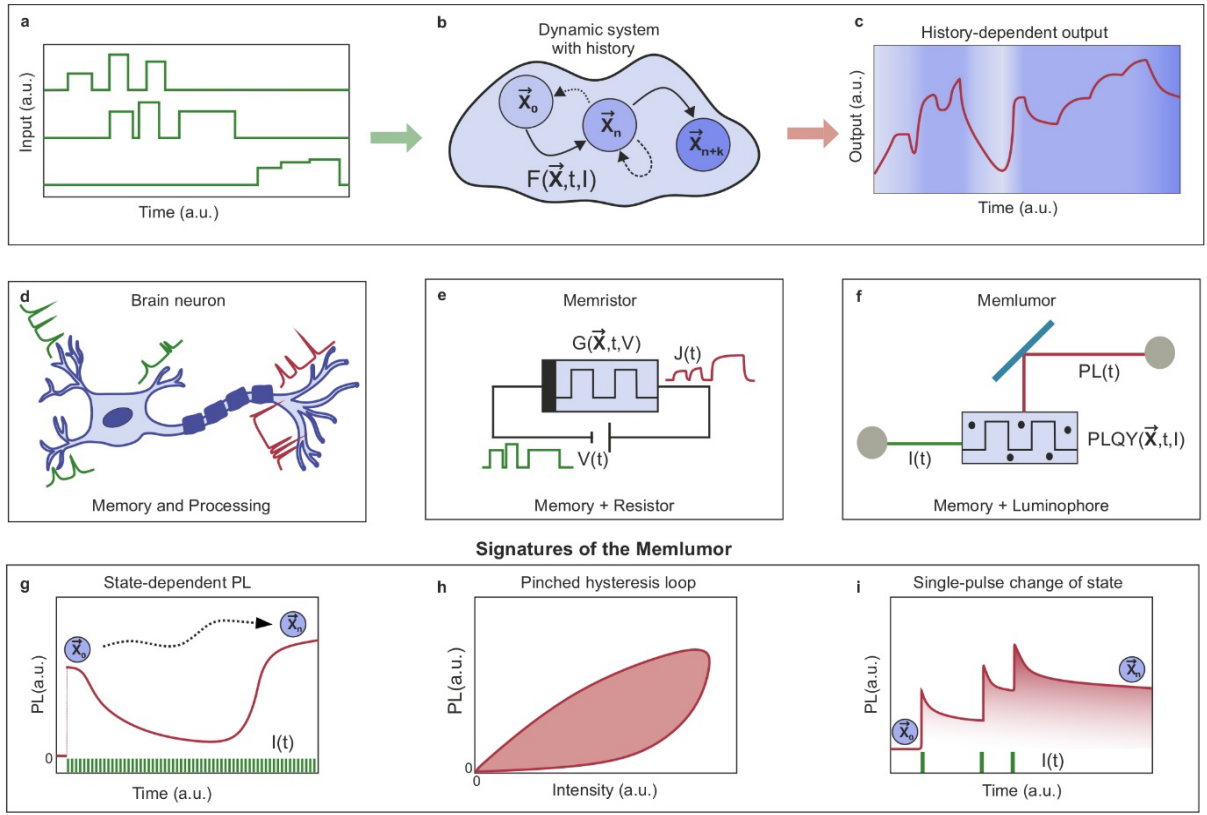

**Fig. 1. From neuron to memristor and memlumor.** Illustration of a complex time-dependent input signal (a) applied to a dynamic system (b) transforming it to a time-dependent output (c). The output depends on the history of the input I(t) via a time-dependent state vector $\vec{X}$ of the system and the transfer function F($\vec{X}$,t,I(t)). The arrows illustrate possible transitions between states. $\vec{X}_0$ illustrates the initial state, $\vec{X}_n$ is one of the states which can reversibly change back to the initial state $\vec{X}_0$ (volatile memory). Transformation $\vec{X}_n \rightarrow \vec{X}_{n+k}$ is irreversible representing permanent memory. Examples of such dynamic systems with history are: brain neuron (d), memristor (e) and memlumor (f). Neuron (d) transforms electrical time-dependent electrical inputs into electrical outputs. The voltage-driven memristor (e) transforms the voltage signal V(t) into the current signal J(t) through a state-dependent conductance function G($\vec{X}$,t,V(t)). The memlumor shows transformation of the light intensity input I(t) to the photoluminescence PL(t) output via the state-dependent photoluminescence quantum yield function PLQY($\vec{X}$,t,I). Memlumor possesses the dynamic state-dependent features: state-dependent PL output under the constant input (g), hysteresis of PL output versus input intensity (h) and the single-pulse change of the state $\vec{X}$ (i).

Inspired by this, we propose a novel concept of a memlumor (Fig. 1f) as an all-optical memristive device with an output (photoluminescence) that depends on the input (light intensity) and a state-dependent PL quantum efficiency. Mathematically the behavior of a memlumor can be formulated as:

$$PL(t) = PLQY(\vec{X}, I, t)\, I(t) \qquad (3)$$

$$\frac{d\vec{X}}{dt} = f(\vec{X}, I), \qquad (4)$$

where *I(t)* is the excitation light intensity (input, equivalent to voltage in memristors) with a certain spectrum (either narrow or broad), and the *PL(t)* is the photoluminescence signal (output, equivalent to current in memristors). $PLQY(\vec{X}, I, t)$ is photoluminescence quantum yield. By means of this definition, $PLQY(\vec{X}, I, t)$ is the full mathematical equivalence of the memristive conductance $G(\vec{X}, V, t)$. Therefore, state-dependent $PLQY(\vec{X}, I, t)$ represents synaptic weight of the memlumors. Notably, considering that PL output response to switching off the input is not immediate (the characteristic time is the PL decay time[33]), the ratio *PL(t)/I(t)* can be equal to infinity. Similarly, this is inherent to real biological neuron-like, electrically-based outputs.[34,35] We note, that the presented formalism allows changes of the state vector $\vec{X}$ to be both reversible (volatile) and irreversible (non-volatile).

The consequence of eq. 3 - 4 is that the PL response can vary over time because it is the function of the time-dependent state $\vec{X}$ of the system (Fig. 1g). In such a case, probing the PL upon scanning the excitation intensity would give rise to hysteresis (Fig. 1h) and a pulsed input if leading to changes in the state vector $\vec{X}$ after each pulse, would result in a complex PL output (Fig. 1i). These examples justify that realization of the memlumor concept is possible with a luminophore whose PL quantum yield is strongly dependent on the specific state $\vec{X}$ of the sample and its history.

Metal halide perovskites (MHPs) are a fascinating class of materials, with excellent optoelectronic properties and impressive performance in functional devices such as solar cells and light-emitting diodes.[36] Despite the many advantageous properties of MHPs, it is well-established that many reversible and irreversible light-induced processes, notably at wide range of timescales, may significantly alter their behavior and function.[37] These processes originate from the rich physics and chemistry of defects in MHPs and include creation, migration and elimination of ionic and electronic defects.[38–40] In the context of the PL of MHPs, these processes lead to the observation of dynamic processes such as photodoping, photobleaching, PL enhancement, light-soaking and self-healing.[41–46] Importantly, the PL of MHPs is also impacted by the interaction of defects with the environment and the history of the sample.[42] The presence of these dynamic processes and the availability of defect-engineering approaches for their control make MHPs the perfect material platform to explore and demonstrate the memlumor concept.

To explore the applicability of MHPs as memlumors we selected a CsPbBr$_3$ perovskite as a model system (**Supplementary Note 1**) and utilized an experimental setup based on a wide-field fluorescence microscope (Fig. 2a and **Supplementary Note 2.1**). In short, the setup involves the excitation of the sample with a 485 nm pulsed laser with 200 ps pulse width either continuously at a desired repetition rate (from <1 Hz to 80 MHz) or via periodic bursts of N pulses with a desired duty cycle. The PL intensity as a function of time is detected either using a charge-coupled device (CCD) camera with up to 100 ms time resolution or by a time correlated single photon counting (TCSPC) method with 0.5 ns time resolution. The setup allows running experiments automatically according to a pre-designed experimental procedure (**Supplementary Note 2**). Therefore, we can track the dynamic changes the state $\vec{X}$ of the sample across a very wide timescale through their impact on the PL output.

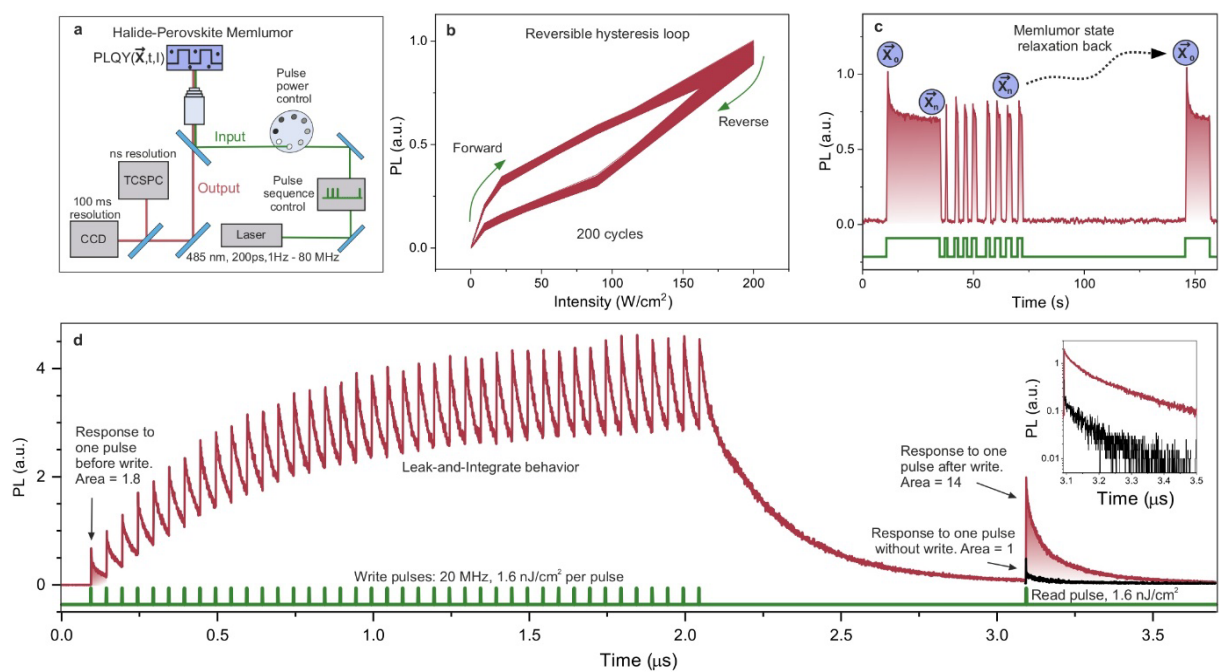

**Fig. 2. Demonstration of the memory effects in the CsPbBr$_3$ thin film memlumor.** (a) Experimental setup. (b) Hysteresis of PL intensity of CsPbBr$_3$ film upon scanning excitation power density, laser pulse repetition rate 80 MHz. (c) Time dependent PL intensity (red) under interrupted excitation (shown by green). (d) Demonstration the of memory effect with a microsecond relaxation time. PL response (red) to a write burst consisted of 40 pulses separated by 50 ns (pulse fluence 1.6 nJ/cm$^2$, the total repetition period of the excitation sequence is 402 µs) followed by one reading pulse after 1 µs delay (laser pulses shown by green). Compare it with the PL response to one single read pulse without any writing bursts (black), see inset in (d). Relative areas under the PL decays excited by one pulse without write burst (area=1), one pulse after the write burst (area=14) and under the PL decay induced by the very first pulse of the write burst are indicated (area=1.8).

To demonstrate the state-dependent nature of PLQY in CsPbBr$_3$, we characterized the PL intensity as a function of the excitation power density measured upon its repeated scanning from 5 mW/cm$^2$ to 160 W/cm$^2$ and backwards (**Supplementary Note 2.2 and 2.3**). The measurements revealed a clear hysteresis loop (**Supplementary Note 3.1**) that was observed for each of the nearly 600 measurement cycles. After a rapid initial evolution of the PL values, the hysteresis loop was stabilized and remained largely unchanged across 200 cycles (Fig. 2b,

**Supplementary Note 3.1**). Such hysteretic behavior is a fingerprint for the presence of volatile light-induced processes that alter the PL at time scales that range from seconds to minutes.

The impact of the excitation history on the state of the memlumor is visualized in Fig. 2c in the time domain. Upon switching on of the excitation light, the initially bright PL (we designate this state of the sample as $\vec{X}_0$) decreases at a sub-second time scale to a lower PL level (we designate this state as $\vec{X}_n$). The transition from $\vec{X}_0$ to $\vec{X}_n$ can be considered as "information writing". The stored information can be "read" shortly afterwards, since the low PL intensity is observed when exciting the sample again for a short period of time (Fig. 2c, **Supplementary Note 3.2 and 3.3**). However, this memory effect disappears once the sample is held without light for several tens of seconds, reversing the memlumor to its original state. This reversible PLQY behavior is a full analogy to the reversible conductance behavior in volatile memristors.[47–49] Notably, tuning of the light input allows for an accurate control the state of the sample and the scale of these long-term memory effects (**Supplementary Note 3.4**).

Together with this rather long living volatile memory effect, there are also other memory effects impacting on the PLQY of $CsPbBr_3$ that can be probed at shorter timescales ranging from nanoseconds to milliseconds. To observe these effects, we excite the sample by a writing burst consisting of N closely spaced laser pulses and monitor the time-resolved PL using the TCSPC method (**Supplementary Note 2.4**). Fig. 2d illustrates that the response to a writing burst with N=40 consists of 40 overlapping PL decays with a progressive increase in their initial PL amplitude after each pulse with a slight decay between the pulses. This process is a consequence of a single-pulse change in the state vector $\vec{X}$ of the sample, demonstrating the so-called 'leak-and-integrate' behavior identified for neurons, electrical and optical memory devices.[20–22,50–53]

The information committed to memory by the writing burst is retained for a characteristic retention time, within which it can be read. In Fig. 2d, the process of reading the stored information is performed by an excitation with a single read pulse 1 μs after the writing burst was completed and the PL signal has decayed. The PL response to the single read pulse is a factor of 14 higher than that obtained without the write burst (areas under the PL decay curves are indicated). This enhancement factor can be varied by changing the number of pulses in the writing burst, making it possible to write and store information in the memlumor. Importantly, this response is the direct evidence of the state-dependent $PLQY$ change induced by the writing burst. Such a response would only be possible in case that certain components of the vector $\vec{X}$ are altered for much longer than the PL decay time. Therefore, memlumors based on $CsPbBr_3$ film are of at least second order having at least two independent state-variables.[27]

While the characteristic retention time is quite short, our results indicate that an additional accumulation effect is also impacting on the behavior of the memlumor. This is evident by the observation that the response to the first pulse of the writing burst is a factor of 1.8 larger than that to a single read pulse without a writing burst. This is a consequence of the fact that the TCSPC experiment is repeated with a cycle of T=402 μs, so that each new writing burst arrives approximately 400 μs after the reading pulse of the previous cycle. The enhanced response to the first pulse in each writing burst suggests that the memlumor's response is impacted by its longer-term history, *i.e.* the previous cycles of excitation. Accounting for this, we can conclude that the factor 14 enhancement of PLQY discussed above is in fact a combination of a factor 7.7 increase due to the writing burst of this cycle and 1.8 times increase due to many previous writing bursts (accumulation effect). It means that varying the repetition period T of the write/read cycle offers the computing opportunity to tune the way information is stored in the memlumors.

Understanding the physical mechanisms that underpin the behavior of memlumors is critical for their future development and applications towards high-order neuromorphic computing. The function of a memlumor is based on the use of a material with a PLQY that is affected by its excitation history. For the memory to exit, there should be something which can store this information in the material. Obviously, these must be components of the state vector $\vec{X}$ and their physical meaning should be identified.

The radiative recombination rate of a semiconductor is proportional to the product n(t)p(t), where n(t) and p(t) are the time-dependent concentrations of free electrons and holes, respectively. For an undoped intrinsic semiconductor, the density of charge carriers in dark is very low, so, essentially all charge carriers are photogenerated by light. However, upon photogeneration, charge carries may become trapped in defect states (Fig.3a, b) and remain in them for a significant time (µs to ms) until they eventually recombine with charges of the opposite sign.[33,54–56] Thus, the traps are able to hold the memory of the photoexcitation history for some time. The presence of traps and their impact on photoluminescence – which is normally seen as disadvantageous in the context of optoelectronic applications – is the fundamental prerequisite that enables the PLQY to be impacted by the material's photoexcitation history (**Supplementary Note 4**). Following the formalism defined in equations (3) and (4), the properties and dynamics of these traps should be included in the internal state vector $\vec{X}$.

To demonstrate how traps impact on the state vector $\vec{X}$, let us assume the simplest case of a material with an efficient trapping of only one type of charge carries, *e.g.* electrons. Upon photoexcitation, a significant number of the generated free electrons becomes trapped ($n_t$), leading to a situation where the concentration of free electrons (n) is much smaller than that of the free holes (p), i.e. p>>n and p=$n_t$+n due charge neutrality. (Fig.3a). This scenario is typically referred to as *photodoping*,[33,39,46,56] and the PL of such a system is impacted by the two independent variables n and $n_t$ that are included in the state vector $\vec{X}$ =(n,$n_t$, … ).

Expanding on this example, one may consider the traps as *memory cells* (Fig.3a,b,c), which can be "written into" by occupying them with electrons, and by doing so, increasing the concentration $n_t$. In this case, if the arrival of each new pulse in a writing burst precedes the recombination of the trapped electrons, their density ($n_t$) – and with it the PL response of the memlumor - will progressively grow with each subsequent writing pulse (Fig.3a and 3b). In this simplest case the memory cells are static and volatile emulating neuronal response of the memlumors (Fig.2d and Fig.3d,e, **Supplementary Note 5 and 6**).

Importantly, the information stored in these static memory cells (i.e. $n_t$) can be read by a weak reading excitation pulse arriving with a substantial delay after the writing burst when the PL intensity is close to zero (Fig. 2d). The short reading pulse generates $n_0$ free electrons and $n_0$ free holes, resulting in a total of $n_0$ electrons, but $n_0+n_t$ holes. The resulting PL intensity just after the pulse arrival is proportional to $n_0(n_0+n_t)$, or, because $n_t \gg n_0$, to $n_0 n_t$, making it proportional to the number of electrons stored in the traps (i.e. filled memory cells).

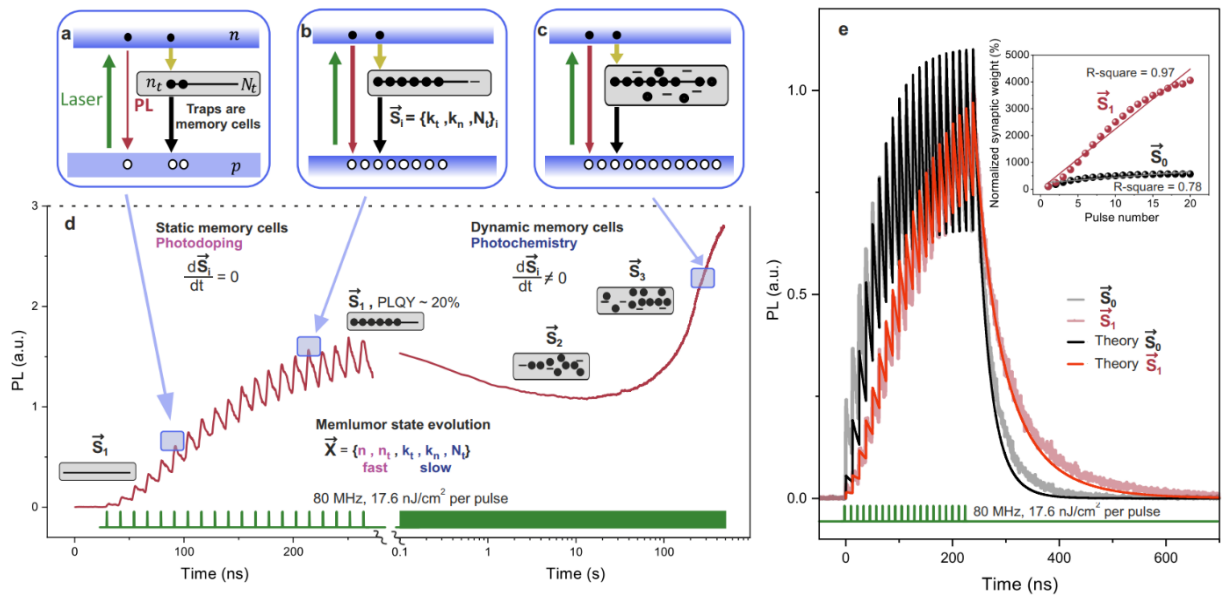

**Fig. 3. Physical processes behind the CsPbBr$_3$ memlumor operation.** a), b) and c) – band diagrams explaining the photodoping effect occurring upon photoexcitation. Each trap can be seen as a memory cell able to capture one electron, from a) to b) the number of trapped electrons increases. Properties of the traps are characterized by the state vector which is the same for a) and b) but different for c). e) Shows the PL response of CsPbBr$_3$ film to 80MHz pulsed excitation (illustrated by green in the bottom) over the time scale from nanoseconds to minutes in connection with the band diagrams a), b) and c). d) The scheme showing four basic steps of working with memlumors. e) PL response to of burst of 20 laser pulses for the same CsPbBr$_3$ film in two different states plotted together with fits obtained from SRH+ model. Inset shows the changes in the linearity of the of the synaptic weight for these different states of the sample.

The dynamics of the processes introduced above can be described by a system of differential equations based on the Shockley-Read-Hall (SRH) charge recombination model with added radiative and Auger recombination (SRH+ model, **Supplementary Notes 7.1, 7.2 and 7.3**) which was introduced in our previous work.[33] To apply this model in practice, it is necessary to extract all the model parameters for the particular sample investigated here. Among them, the most important parameters are the rate constants of radiative recombination ($k_r$), trapping ($k_t$), non-radiative recombination of trapped electrons with free holes ($k_n$) and the trap concentration ($N_t$). To quantify these parameters, we utilized a recently developed method[33,57] that maps the PLQY of a sample over a large range of pulse repetition rates (f) and pulse fluences (P) **(Supplementary Notes 2.5 and 4)**. Note that PLQY(f,P) dependence is the dependence of the memlumor's synaptic weight on f and P. These PLQY mapping measurements are complemented with PL decay kinetic measurements (**Supplementary Notes 7.5 and 7.6**). The data from these experiments was fitted using the SRH+ model and the obtained model parameters are summarized in **Supplementary Table 6**.

Using the parameters obtained for the CsPbBr$_3$ film as the starting values it is possible to semi-quantitatively reproduce the 'leak-and-integrate' experiment presented in Fig. 2d (**Supplementary Note 7.6, Supplementary Fig.16, Supplementary Table 9**) by solving the full set of differential equations. Therefore, the processes described by the SRH+ model naturally explain the memory effects observed on the fast timescale, where the state vector $\vec{X}$=(n,n$_t$,...) changes with each individual excitation pulse.

Importantly, while the SRH+ model is descriptive of the memlumor's behavior on fast time scales, it predicts that a quasi-steady state constant PL should be reached rather rapidly, typically within a few microseconds (**Supplementary Fig. 13**). On the other hand, our experimental results (Fig. 2b and 2c) indicate that this is not the case, suggesting that the state vector $\vec{X}$ evolves on longer timescales (seconds), *i.e.* exhibiting a longer-term memory.

The observation of this longer-term memory behavior makes is necessary to expand the model introduced above to also include the dynamics of *memory cells*, *i.e.* introduce a time dependence to the memory cell parameters, i.e. k$_t$(t), k$_n$(t) and N$_t$(t). This can occur as a consequence of long-term photochemical processes that lead to the evolution of defect properties.[39,42–44,58] It is thus convenient to define the vector $\vec{S}$ to represent the dynamic memory cell parameters: $\vec{S}$ = (k$_n$, k$_t$, N$_t$). Hence, the full state vector $\vec{X}$ contains both fast (n,n$_t$) and slow (k$_n$, k$_t$, N$_t$) changing variables, i.e. $\vec{X}$ = (n,n$_t$, $\vec{S}$, ...). Notably, tuning of $\vec{S}$ vector needs much more energy than a photodoping process (which is theoretically as low as an energy of a single photon) (**Supplementary Notes 2.5).** However, when changed, the new state of $\vec{S}$ vector can exist for timescale for at least several days. The combined picture of static and dynamic memory cells active across different timescales, makes it possible to explain the PL(t) dependence shown in Fig. 3d across the entire timeframe from ns to minutes. To summarize, the trap evolution under applied optical stimuli explains the coexistence of both volatile and non-volatile memories across a wide range of times and energies in a memlumor.

Obviously, the parameters of vector $\vec{S}$ determine not only the equilibrated value of PL at long timescales, but also the response of the system to input pulses at the nanosecond regime that also depends on k$_t$, k$_n$ and N$_t$. This is illustrated in Fig.3e where the response of the sample to a writing burst of 20 pulses differs between a sample at an initial state $\vec{S_0}$ and that in a state $\vec{S_1}$, which is achieved by pre-exposing it to high intensity light illumination (bleaching). Notably, $\vec{S_1}$ state has a non-volatile nature and its properties permanently changed for at least several days (see further Fig. 5c). Both types of memlumor states ($\vec{S_0}$ and $\vec{S_1}$) responses can be well simulated theoretically by changing the vector $\vec{S}$ in the SRH+ model (**Supplementary Notes 7.5 and 7.6, Supplementary Tables 7 and 8**). Importantly, considering that the PLQY of a memlumor is equivalent to the conductance (synaptic weight) of a memristor, these results show that varying the vector $\vec{S}$ makes possible a non-volatile tuning of both the single-pulse like linearity and absolute value of the memlumor synaptic weight (see inset in Fig.3e) emulating the function of synapses in brain.

The realization of the memlumor concept offers a broad range of possibilities for its integration into neural networks due to the wealth of potential methods to address, couple and read from these all-optical devices. For example, it is possible to employ them using free space coupling (Fig. 4a), coupling by waveguides (Fig. 4b,c) and coupling via near field. In the free-space regime, the excitation (input) and emission (output) light propagate in free space and can be delivered usings optical elements (e.g. lenses) to a desired location with an accuracy determined by the light diffraction limit. In practice, such a design would lead to the limitation in the smallest addressable volume (size of an active element) to be approximately 1 cubic micron (Fig. 4a). Note that the memlumor itself can be smaller, however, to address individual memlumors independently the distance between them should be larger than the diffraction limit. Importantly, in such a scenario, there would be no need for wiring or interconnects, since the connections are fully optical.

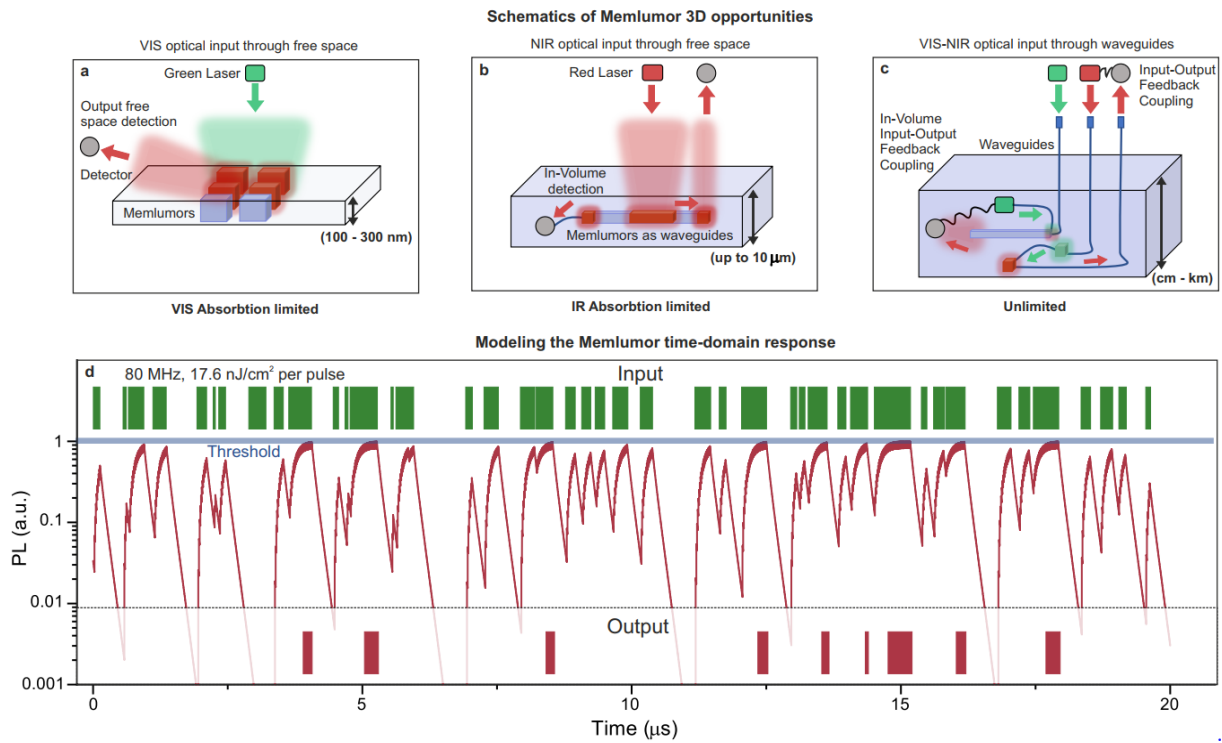

**Fig. 4. Possible applications of memlumors for in-memory calculations and neural networks.** a) 2D device with in-plane arrangements of individual memlumors addressed in far field by free-space optics. b) quasi-3D device combining free-space input using 2-photon excitation, internal connections by waveguides and output via free space to an external detector or via waveguiding to an internal detector. c) Hypothetical 3D memlumor device using internal and external detectors coupled by waveguiding. d) Calculated PL response (red) to the input signal presented by green of the memlumor with the model parameters obtained for $CsPbBr_3$ film (**Supplementary Table 6**). The output is not zero if the PL signal exceeds the threshold (grey line). Memlumor converts the green barcode to the red barcode.

Moreover, for certain applications, it is not necessary to be able to read each individual memlumor, and instead the total output of a certain group of devices (e.g. one row of a cross bar for matrix multiplication[15], or all memlumors together for summation operations) is of importance. These operations are natural, because the bosonic nature of photons allows the summation of the outputs from independent memlumors towards the same optical output detecting channel without their mutual interference. This promises wide parallelism in computing operations and the ability to continuously process online-updated data without information losses.

Another opportunity with memlumor systems is using the wavelength spectrum to tune the optical neuromorphic system response. For example, the fact that the PL energy is slightly red-shifted due to the Stokes shift opens additional opportunities for the integration of memlumors into three-dimensional (3D) architectures. This is because the lower energy PL photons can propagate inside the semiconductor over a long distance without substantial attenuation. Thus, a fraction of the PL light generated in the depth of a thick crystal can still be detected by an external photodetector. This lifts the requirement that the devices have to be two dimensional (2D) in nature (Fig.4b). Importantly, it is possible to address memlumors using a two-photon excitation with near infrared (NIR) laser, leading to a spatial resolution of 2 micrometers or better in the depth dimension (Fig.4b). We envision that a confocal microscope setup with one and two-photon excitation capabilities and the point (confocal) and wide field excitation and detection modes is an ideal instrument for demonstrating the possibility to exploit memlumors for neuromorphic computing.

An alternative approach to addressing, reading and coupling memlumors is based on the use of waveguiding (Fig.4b,c). MHPs exhibit high refractive indices making them excellent waveguiding materials. This makes it possible to combine the functionalities of memlumors with those of waveguides in the same material platform.

By showing such diversity of possible 2D and 3D integration, we, next, vision spiking neural networks as the most perspective neural network architecture for the realization of memlumor-based photonic neuromorphic computing in the nearest future. In fact, having a wide time range of volatile physical mechanisms supported by non-volatile memory, makes memlumors ideal material platform for such purpose. To illustrate the suitability of memlumors for the realization of spiking neural networks (SNN), we utilized the SRH+ model parameters extracted from the experimental data (**Supplementary Table 6**) to simulate the transformation of a time-dependent input to an output 'barcode' by applying a detection threshold (Fig. 4d).

It is important to highlight that the practical realization of systems based on memlumors do not require the costly TCSPC setup utilized here solely in order to investigate their photophysics. The input can be generated utilizing inexpensive commercial pulsed laser diodes or even light-emitting diodes (LEDs) that are capable of generating the desired pulse burst sequences in the time range from nanoseconds to microseconds with an amplitude modulation, while narrowband photodetectors could be used for detecting optical output. These requirements are already fully satisfied in standard optical communication systems, where the bandwidths in the range up to tens of GHz are standard.

Some application ideas presented above can be realized experimentally using $CsPbBr_3$ microcrystals as memlumors. For example, by coupling three $CsPbBr_3$ microcrystals on a GaP waveguide (**Supplementary Note 1**),[59] we are able to realize a summation operation (Fig 5a, **Supplementary Note 8**). Upon excitation, the response of all three memlumors is summed and detected at the edge of the waveguide. Another experimental demonstration relies on the use of a CsPbBr$_3$ microwire that serves both as a memlumor and as a waveguide (Fig. 5b). In this case, it is possible to tune the synaptic weight by changing the laser excitation position on the microwire.

To examine the competitive potential of memlumors, we characterized their different figures of merit focusing on the CsPbBr$_3$ perovskite model system. We first performed the non-volatile switching of the memlumor (applying maximum intensity of 160 W/cm$^2$) from initial state $\vec{S}_0$ to the state $\vec{S}_1$ similar to that illustrated in Fig. 3e, which required a total energy of less than 1 $\mu J$. Next, we applied a repetitive 20 pulse-burst with a total energy of <3.5 fJ per period (equivalent to ~300 emitted photons per one burst estimated for PLQY=20% which is a realistic number for CsPbBr$_3$ (**Supplementary Fig.8 b**)). The stability of $\vec{S}_1$ state together with repetitive threshold-defined volatile switching (a switching time of ~250 ns) has been evaluated continuously for 100 hours (Fig. 5c). The outcome of this long-term measurement is that the memlumor's total endurance is better than $10^{11}$ pulses, corresponding to $10^{10}$ cycling operations. The stability of the threshold switching remained largely unchanged with variations of about 20%. Note that the sample was kept at ambient air without any encapsulation during these studies, exemplifying its high stability. These figures of merit demonstrate the enormous potential of memlumors for the realization of photonic neuromorphic computing with a spiking neural network architecture combining both volatile and nonvolatile memory in one photonics hardware.

Addressing the potential long-term stability of memlumors, it is important to highlight that unlike perovskite based solar cells, memlumors are exposed to much lower light intensities (only 5.6 mW/cm$^2$ time-average power density in our experiments) and their input does not contain the particularly harmful ultraviolet (UV) spectral range. Moreover, memlumors do not have charge extracting layers and electrical contacts, so they do not suffer from ion migration induced degradation of these additional layers that is one of the main causes of performance loss in solar cells.[60,61] Considering these differences, and the fact that perovskite-based solar cells have already reached one year stability under continuous illumination at 100 mW/cm$^2$ average power,[62,63] it is reasonable to expect that the long-term stability of memlumors will significantly surpass that of perovskite solar cells.

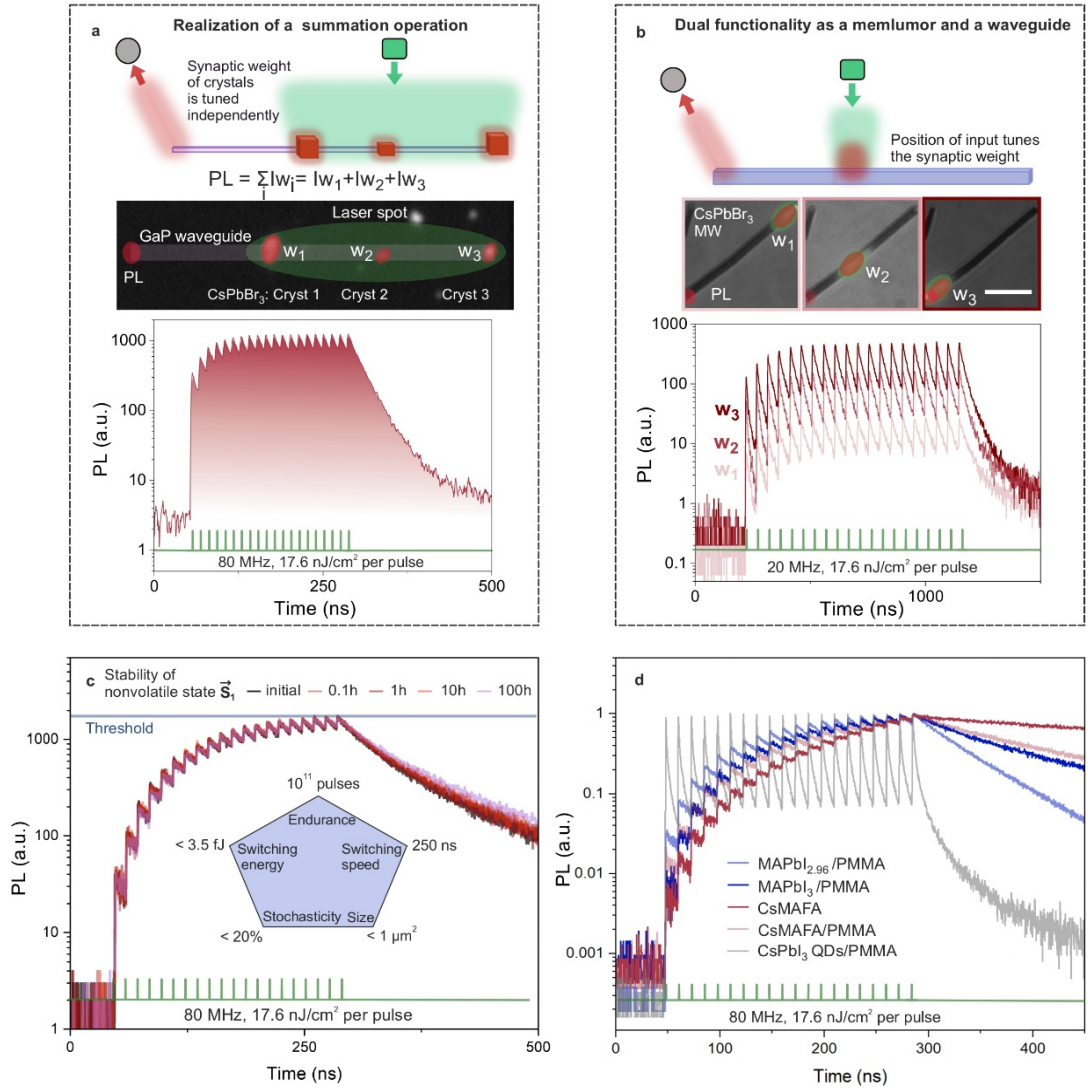

**Fig. 5.** a) Summation of the PL signal from three CsPbBr$_3$ crystals coupled to a GaP waveguide. PL is detected from the end of the waveguide. Top panel - schematic, middle panel – PL image of the structure with the PL from the crystal colored by red. The bottom panel: time-resolved PL signal collected from the end of the waveguide when excited by a burst 20 pulses. b) Using a CsPbBr$_3$ microwire as a spatially extended memlumor and the waveguide at the same time. Top panel – schematic, middle panel – PL images of the CsPbBr$_3$ wire under three different locations of the focused excitation. Bottom – time-resolved PL signal collected from the end of the wire (marked by a red spot in the image) for the three positions of the excitation beam. c) Volatile response of the memlumor in the nonvolatile state $\vec{S}_1$ monitored continuously for 100h (~$10^{10}$ write/erase cycles). The inset shows several figures of merit of CsPbBr$_3$ memlumor with volatile $\vec{X}$ state dependent on nonvolatile state $\vec{S}_1$. d) The normalized leak-and-integrate behavior (PL response to 20 excitation pulses) measured for different metal-halide perovskite samples.

The versatility of MHPs material platform for the realization of the memlumor concept was tested by examining the leak-and-integrate behavior of perovskite samples with different composition and dimensionality (Fig. 5d). In the case of both methylammonium triiodide (MAPbI$_3$) and triple cation (CsMAFA) perovskites, we observe a clear leak-and-integrate behavior in both cases but observe that the exact choice of composition plays an important role in determining the exact dynamics of the response. Moreover, the response can be further modified by either defect engineering or interfacial modification. The former is illustrated for MAPbI$_3$, where a vacancy-rich sample was prepared by lowering the stoichiometry to 2.96,[57,64] leading to a slightly different leak-and-integrate behavior. The impact of interfacial modification using poly(methyl methacrylate) (PMMA) as an example is illustrated for the CsMAFA case. Importantly, the leak-and-integrate behavior is not ubiquitous to all perovskite samples. For example, for the same excitation conditions we did not observe the leak-and-integrate behavior in the case of CsPbI$_3$ perovskite quantum dots. All this means that appropriate chemical engineering and miscellaneous material synthesis is needed towards creating memlumor neuromorphic platforms.

To conclude, we introduce memlumors as a new type of an all-optical dynamic memory device that combines a luminophore and memory. The function of a memlumor is formalized as a full mathematical equivalence to that of memristors, with state-dependent $PLQY(\vec{X}, I, t)$ being fully analogous to state-dependent conductance $G(\vec{X}, V, t)$ in memristors. We demonstrate the memlumor concept using MHPs as a material platform that makes it possible to exploit its rich defect physics to have a direct impact on the memlumor's properties. Specifically, the combination of photodoping and photochemistry phenomena makes it possible to utilize traps in perovskite semiconductors as static and dynamic memory cells that define both volatile and non-volatile memory effects on timescales ranged from ns to days.

High PLQY of MHPs, their compositional versatility, and the existence of a broad range of material engineering approaches for their control, make MHPs suitable for the application-driven property control of the memlumors. Importantly, the general concept of the memlumor with a state-dependent $PLQY(\vec{X}, I, t)$ can be extended beyond MHPs, since other material classes that exhibit similar photophysical phenomena can also be explored. For example, photodoping has been observed, among others, in organic materials, 2D materials and metal oxides making them promising candidates for future investigation.[65–67]

By elucidating their fundamental photophysical principles, we demonstrate that the description of memlumors using the Shockley-Read-Hall-based model suggests their state vector $\vec{X}$ to contain both fast (n, n$_t$) and slow (k$_n$, k$_t$, N$_t$) changing variables existing in a broad range of timescales and light input energies. This makes memlumors particularly promising for further integration and application in developing neuromorphic computing using spiking neural network architectures.

Importantly, the maximum value of the state-dependent PLQY can be up to 100% and still be controlled by population of the defect states ensuring an acceptable level of losses in optical schemes based on the memlumors. Even in our standard CsPbBr$_3$ films this value is around 20%.

An initial assessment suggests that the figures of merit of memlumors are highly competitive in comparison to those of other memory technologies in terms of both energy efficiency and stability. Notably, the hardware required for the development of memlumor-based technologies is readily available since it is already commonly used in optical communication systems.

Finally, while our work provides the conceptual framework of memlumors, examines their fundamental operational physical mechanisms and offers several examples for their applications, significantly more research is required in order to realize memlumor-based neural networks. This includes the large-scale system integration of memlumors based on either near- or/and far-field optics, the development of novel algorithms for energy-efficient computing and the realization of on-chip and 3D integrated memlumors in existing photonics neuromorphic components and combining with other neuromorphic photonics technologies such as *e.g.* phase change memories. These examples illustrate how the novel concept of a memlumor will drive innovation in a range of research fields working together towards a gradual transition into next-generation neuromorphic computing paradigm.

**Acknowledgement:**

The work was supported Swedish Research Council (2020-03530). JK thanks Wenner-Gren foundation for the postdoctoral scholarship (UPD2022-0132). This project has received funding from the European Research Council (ERC) under the European Union's Horizon 2020 research and innovation programme (ERC Grant Agreement n 714067, ENERGYMAPS) and the Deutsche Forschungsgemeinschaft (DFG) in the framework of the Special Priority Program (SPP 2196) project PERFECT PVs (#424216076). Y.V. thanks the DFG for the generous support within the framework of the GRK 2767 (project A7).

# Supplementary information for

# Memlumor: a luminescent memory device for photonic neuromorphic computing

Alexandr Marunchenko[1*], Jitendra Kumar[1], Alexander Kiligaridis[1], Shraddha M. Rao[1], Dmitry Tatarinov[2], Ivan Matchenya[2], Elizaveta Sapozhnikova[2], Ran Ji[3,4], Oscar Telschow[3,4], Julius Brunner[3,4], Anatoly Pushkarev[2], Yana Vaynzof[3,4], Ivan G. Scheblykin[1*]

*Corresponding author(s). E-mail(s):shprotista@gmail.com, ivan.scheblykin@chemphys.lu.se;

[1] Chemical Physics and NanoLund, Lund University, P.O. Box 124, 22100 Lund, Sweden

[2] School of Physics and Engineering, ITMO University, 49 Kronverksky, St. Petersburg 197101, Russian Federation

[3] Chair for Emerging Electronic Technologies, Technical University of Dresden, Nöthnitzer Str. 61, 01187 Dresden, Germany

[4] Leibniz-Institute for Solid State and Materials Research Dresden, Helmholtzstraße 20, 01069 Dresden, Germany

# Table of content

# Supplementary Note 1. Synthesis of Materials for Memlumors

## 1.1 CsPbBr$_3$ perovskites

*Materials:*
Cesium bromide (CsBr, 99.99%, TCI Chemicals), cesium carbonate (Cs$_2$CO$_3$, 99%, Sigma-Aldrich), lead(II) bromide (PbBr$_2$, 99.999% trace metals basis, TCI chemicals), dimethyl sulfoxide (DMSO, anhydrous ≥99.8 %, Sigma-Aldrich), diphenyl ether (DE, ≥99%, Sigma-Aldrich), oleic acid (OA, technical grade, 90%, Sigma-Aldrich), oleylamine (OLAm, technical grade, 70%, Sigma-Aldrich) were used as received.

*Preparation of perovskite precursor solution:*
CsBr (62 mg) and PbBr$_2$ (110 mg) were mixed and dissolved in DMSO (1 mL) by shaking without heating to obtain a clear 0.3M solution. The solution was filtered through a 0.45 μm PTFE syringe adapter right before the procedure of perovskite film deposition.

*CsPbBr$_3$ polycrystalline film preparation:*
The glass substrates (15 x 15 mm$^2$) were cleaned by sonication in NaHCO$_3$ solution, deionized water, acetone, and 2-propanol for 10 min consecutively, and then exposed to UV-generated ozone for 15 minutes to obtain a hydrophilic surface. Afterwards, substrates were transferred in the dry glovebox filled with N$_2$ gas. The deposition of perovskite films was conducted on the substrates by single-step spin-coating method at 3000 rpm for 5 minutes. Thereafter, the samples were gradually annealed on a hot plate from 50 °C up to 130 °C for 15 min to remove dimethyl sulfoxide residues and complete the crystallization of perovskite.

*CsPbBr$_3$ microwires preparation:*
For the CsPbBr$_3$ microwires (MWs) synthesis, the temperature difference-triggered growth method was used[1]. For this, a furnace (PZ 28-3TD High-Temperature Titanium Hotplate and Program Regler PR5-3T) was employed to control the temperature during the MWs growth. The CsPbBr$_3$ perovskite material was sublimated from a source substrate to a target substrate. As a source substrate, the CsPbBr$_3$ polycrystalline film (prepared by the method described above) was used. The target sapphire substrate (10 x 10 mm$^2$) was cleaned by sonication in deionized water, acetone, and 2-propanol for 10 min consecutively. Two substrates were separated with an air gap of 0.5 cm. The temperature of both substrates was controlled by the furnace temperature. The synthesis was initiated at a furnace temperature of 350 °C. Afterwards, the temperature was increased up to 520 °C for 10 min and kept unchanged for 10 min. As a result, the sublimated CsPbBr$_3$ microwires were grown in three crystallographic directions of the saphite.

*CsPbBr$_3$ NPs on GaP nanowaveguides preparation:*
Cs$_2$CO$_3$ (0.407 g) was loaded into a 100 mL flask along with octadecene (20 mL) and oleic acid (OA, 1.25 mL), dried for 1 h at 120 °C, and then heated up to 150 °C to give a clear CsOA 0.125M solution. All manipulations were conducted in a N$_2$-filled glove box. Perovskite nanoparticles (NPs) were synthesized using a Schlenk line by modified hot-injection method. PbBr$_2$ (35 mg) was added to a 50 ml two-neck flask and dried in DE solvent at 120 °C under vacuum for 40 min. Then, OA (150 μL) and OLAm (150 μL) were added dropwise to it. Thereafter, the flask was purged with N$_2$ and the mixture was heated up to 150 °C to obtain a clear solution. The solution was cooled down to 88 °C before the injection of preheated at 120 °C CsOA solution (0.5 mL) followed by incubation for 2 h at 88 °C. Afterwards, the solution was heated up to 150 °C, incubated for 30 min, and, finally, quenched by using an ice bath. The product was centrifuged at 3000 rpm for 5 min, separated from supernatant solution, and redispersed in 10 mL of n-hexane. Large particles were settled down in 5 min and upper fraction of the solution (3 mL) was pipetted for further manipulations. This fraction contains a wide dispersion of NPs with the mean size of about 360 nm according to dynamic laser scattering measurements.

GaP NWs were grown on a Si substrate by molecular beam epitaxy using a protocol reported by Trofimov et al. [2] NWs were transferred into suspension by ultrasonication of 0.5x0.5 cm substrate in 1 mL of 2-propanol, drop-casted onto a glass substrate, and rinsed with hot acetone two times. Then, the solution containing $CsPbBr_3$ NPs was drop-casted over GaP NWs to give nanowaveguides decorated with subwavelength light emitters.

## 1.2 Other metal-halide perovskites

*$MAPbI_3$ and $MAPbI_{2.96}$ films:*
MAI and $Pb(Ac)_2·3(H_2O)$ (at 3:1 or 2.96:1 molar ratio) were dissolved in anhydrous *N,N*-dimethylformamide (DMF) with a concentration of 40 wt% with the addition of hypophosphorous acid solution (6 μL / 1 mL DMF). The perovskite solution was spin coated at 2000 rpm for 60 s in a drybox (RH < 0.5 %). After spin coating, the samples were dried for 20 seconds by a stream of dry air. Afterwards, the samples were kept at room temperature for 5 min and subsequently annealed at 100 °C for 5 min.

*Triple cation perovskite films:*
The perovskite precursor solution was prepared by dissolving $PbI_2$ and $PbBr_2$ in a solvent mixture (DMF/DMSO = 4/1) and CsI in DMSO at 180 °C for 10 minutes. After cooling down, CsI, $PbI_2$ and $PbBr_2$ solutions were mixed in a volume ratio of 0.05:0.85:0.15, to obtain an inorganic stock solution. MAI and FAI powders in separate vials were then dissolved to form stock solutions. Finally, the MAI-stock-solution was mixed with the FAI-stock-solution in a 1:5 volume ratio to acquire the final 1.2 M $Cs_{0.05}(FA_{5/6}MA_{1/6})_{0.95}Pb(I_{0.9}Br_{0.1})_3$ perovskite solution in DMF and DMSO. The perovskite layer was deposited via a two-step spin-coating procedure with 1000 rpm for 12 s and 5000 rpm for 28 s. Before spinning, 40 μL of perovskite precursor solution were applied to the sample statically and 150 μL of Trifluorotoluene (TFT) was dripped on the spinning substrate, 5 s before the end of the second spin-coating step in a rapid fashion, in agreement with Taylor et al. [3] Subsequently, the samples were annealed at 100 °C for 30 min.

*$CsPbI_3$ QDs perovskite films:*
The synthesis of the perovskite QDs was done via the hot injection method, firstly published by Protescu et al., with several adjustments [4] Before the synthesis, oleic acid (OA, tech. grade 90%, Sigma Aldrich) and oleylamine (OLA, tech. grade 70 %, Sigma Aldrich) were degassed at 100 °C for 1 h to guarantee high purity of the reactants. Cs-oleate was produced by combining Cs-carbonate with OA. For the preparation of Cs-oleate solution, 0.407 g of $Cs_2CO_3$ (TCI, >98 %), 20 mL of octadecene (ODE, tech. grade 90 %, Acros Organics), and 1.25 mL of OA were loaded in a 2-neck round-bottom flask and degassed for 1 h at 100 °C at vacuum. Thereafter, the flask was filled with nitrogen and heated to 150 °C until all reactants reacted and a clear solution of Cs-oleate was obtained. The Cs-oleate was then stored in nitrogen at 70 °C until usage. For the synthesis of the $CsPbI_3$, 1 g of $PbI_2$ (99.99 %, TCI) and 60 ml of ODE were filled in a 2-neck round-bottom flask and degassed for 1 h at 120 °C in a vacuum. Subsequently, the flask was filled with nitrogen and 6 mL of OLA and 6 ml of OA were mixed in a vial and then injected. The flask was again pumped to vacuum for 30 min, until a yellow transparent solution was obtained. Then, the flask was filled with nitrogen and heated to 170 °C. At the target temperature, 4 mL of Cs-oleate was quickly injected into the flask. The solution turned dark red and after 5 s the reaction was quenched with an ice-water bath. For the purification of the as-prepared $CsPbI_3$ nanocrystals, 12.5 mL of the crude solution was mixed with 37.5 mL of Methyl acetate (MeOAc, 99 %, Acros Organics) and centrifuged for 10 min at 6000 rpm. The supernatant was discarded and the wet $CsPbI_3$ pellets were redispersed in 3 mL of hexane (97%, Acros Organics). The solution was again mixed with 5 mL of MeOAc and centrifuged for 10 min at 6000 rpm. The supernatant was removed, and the precipitates of all tubes were combined in one and dispersed in 25 mL of hexane. This solution was centrifuged for 5 min at 4000 rpm and this time the supernatant was collected and stored overnight at 4 °C. After that, the solution was centrifuged again for 5 min at 4000 rpm. Finally, the supernatant was collected and dried by using a rotary evaporator. The obtained $CsPbI_3$ nanocrystals were dispersed in octane (99%, Acros Organics) at a concentration of 75 mg/mL for further use.

Then, for fabrication of $CsPbI_3$ quantum dot samples we followed the procedure reported previously [5] with slight modifications. Glass substrates were cleaned thoroughly by sonicating them in acetone and isopropanol for 15 min respectively. To activate the surface and remove organic remains the samples were subjected to oxygen plasma for 10 min. After that, they were transferred to a nitrogen-filled glovebox for the deposition of the perovskite quantum dots. The quantum dot solution that was obtained from the synthesis was then spin-coated dynamically on the glass substrates at a speed of 1000 rpm for the first 5 s and 2000 rpm for the following 10 s. To exchange the long organic ligands, the film was in immersed in a ligand solution of sodium acetate (NaOAc, 99.995 %, Sigma Aldrich) in MeOAc for 5 s and was spin-dried, followed by soaking and spin-drying the film in MeOAc for 5 s twice. This procedure was repeated four times to obtain a layer of approximately 250 nm. After that the sample was soaked in a solution of phenethylammonium iodide (PEAI, Great Cell Solar) in ethyl acetate for 10 s and spin-dryed. The PEAI treated sample was then washed with ethyl acetate.

*PMMA layer deposition:*
The PMMA coating of all different types of perovskite films was performed by preparing a solution of polymethyl methacrylate (PMMA, Great Cell Solar) in chlorobenzene (10 mg/ml) that was then spin-coated at 3000 rpm for 30 s on top of each of the layers.

# Supplementary Note 2.
# Optical setup and Methods for Memlumor Measurements

## S2.1 Optical setup

The photoluminescence (PL) was imaged using a home-built wide-field photoluminescence microscope based on Olympus IX71 with 40X (NA=0.6) dry objective lens and an EMCCD Camera (ProEM, Princeton Instruments) as the primary detector. The PL was excited by a 485 nm diode laser (PicoQuant, ca 100 ps pulse width) through an objective lens. The laser was controlled by the laser driver SEPIA LD828 (PicoQuant). This allows to drive the laser at any repetition rate from 80 MHz to 0.2 Hz as well as in a pulse burst mode and even more complex time patterns. The setup was equipped with a shutter and several computer-controlled neutral density filter wheels to adjust the pulse fluence ($P_i$) of the laser and to attenuate the PL light sent to the detectors. In addition to detection by a camera, the setup made it possible to divert the PL signal to the hybrid photomultiplier detector (HPD, Picoquant PMA Hybrid-42) connected to a time correlated single photon counting (TCSPC) module (Picoquant, Picoharp 300) for measurements of PL decays kinetics. The instrumental response function width was approximately 200 ps.

All components of the setup are controlled by a home-designed Labview software. This software allows to execute a designed sequence of measurements (recording a PL image with the camera or a PL decay curve) according to a list of instructions (table with experimental parameters like filters, exposure time, laser repetition rate and so on). This automatic measurement regime allows to not only perform a single complex measurement (like PLQY(f,P) mapping, see **Supplementary Note 4 and 2.5**), but also to repeat the same complex sequence of measurements multiple times (interactions of the sample with light, or excitation history, **Supplementary Notes 2.3 and 3.1**).

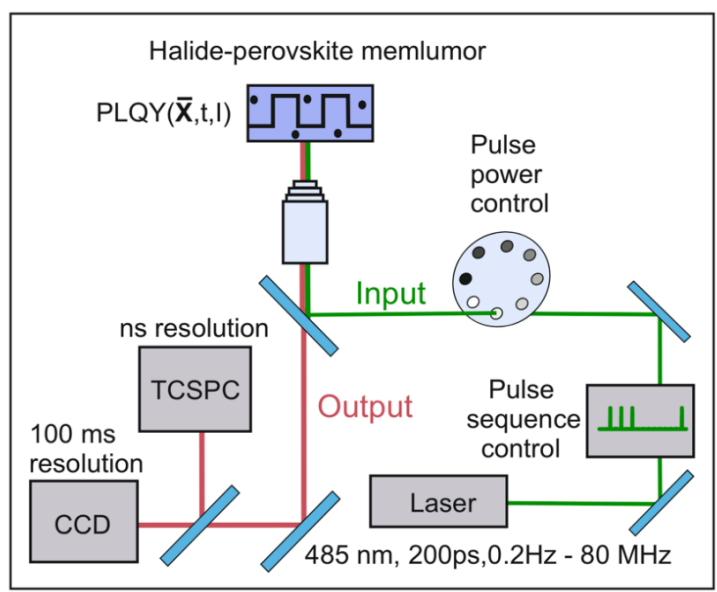

**Supplementary Fig. 1.** Experimental setup used for characterization of memlumors.

## 2.2 Light Excitation Conditions

The initial value of the photogenerated charge carrier density $n_0$ used in the modelling can be estimated for each of the pulse fluence values $P_i$ as $n_0 = \frac{P_i \cdot Abs}{d}$, where d is the film thickness and Abs in the film absorption. $Abs = (1 - S)(1 - R)(1 - e^{-\alpha d})$, where scattering S<<1, R is reflectance and $\alpha$ is the absorption coefficient of the material. In the simplest case $R = \frac{(n-1)^2}{(n+1)^2}$, where n is refractive index of a material at a corresponding wavelength.

For the CsPbBr$_3$ polycrystalline thin film we consider $n \approx 2.3$, $\alpha \approx 0.45 \cdot 10^5 \, cm^{-1}$ at a 485 nm wavelength [6,7]. Thus, for film thickness of 90 nm, Abs=0.28. Therefore, $n_0$ for each of the pulse fluence $P_i$ is equal to:

$$n_0 = P_i \cdot 0.31 \cdot 10^5 \, cm^{-1}$$

The calculated values for $n_0$ are listed in **Supplementary Table 1** below.

**Supplementary Table 1.** Values of the five different pulse fluences P1-P5 mainly used in the experiments, the corresponding charge carrier densities $n_0$, pulse energy densities, and average power densities. The color scheme shown in the table is used in all figures of the paper.

| Pulse fluence $P_i$ (photons/cm²) | Charge carrier density $n_0$ (cm$^{-3}$) | Pulse energy density, (nJ/cm²) | Average power density at 80 MHz repetition rate (W/cm²) |
|---|---|---|---|
| P5 = 4.9x10$^{12}$ | 1.52x10$^{17}$ | 2000 | 160 |
| P4 = 5.4x10$^{11}$ | 1.67x10$^{16}$ | 220 | 17.6 |
| P3 = 4.3x10$^{10}$ | 1.33x10$^{15}$ | 17.63 | 1.41 |
| P2 = 4.0x10$^{9}$ | 1.24x10$^{14}$ | 1.64 | 0.131 |
| P1 = 3.4x10$^{8}$ | 1.05x10$^{13}$ | 0.14 | 0.0113 |

## 2.3. Measurement of PL vs intensity hysteresis

Similar to memristors, where current flow changes the state of the sample, the state of a memlumor changes by exposure to light, leading to a hysteretic behavior of the PL. To record it, the automated measurement setup allows to scan the laser excitation intensity (I) while acquiring the PL intensity of a memlumor, which is fully analogous to scanning the current while recording the voltage for a memristor. To perform such a measurement, a table in a text format (see an example in **Supplementary Table 2**) is uploaded to a Labview software with a set of parameters which then are processed by the program to run the setup according to the required sequence of measurements.

An important consideration while carrying out the hysteresis measurements is not only to choose the appropriate excitation conditions (pulse fluence and pulse repetition frequency), but also adjust the photon flow arriving to the detector. It is important to make sure that the light does not saturate the CCD camera when the excitation intensity is large, and that the signal-to-noise ratio is good enough when the excitation intensity is small. This is ensured by choosing a suitable exposure time and placing a neutral density filter in front of the camera. Notably, the nonlinearity of PL versus I plots is typical for halide-perovskite semiconductors [8,9], and requires the accurate control of the PL intensity in order to satisfy these conditions. To record the PL *vs* I curves of the memlumor, five parameters form the rows in the table in **Supplementary Table 2**. So, each row is one data point of the curve. One additional hidden parameter is the time between two experimental points. It is defined by the speed of the electronics controlling optical density filters. In our case, the time-difference between finishing the acquisition of one data point and starting the next one was around 3 seconds. The table is the record of the excitation history of the sample.

**Supplementary Table 2.** A typical table for the measurement of PL vs I in the scanning regime. During the single cycle the intensity of the excitation power density (pulse fluence $P_i$ multiplied by a constant frequency 80MHz) first increases from the minimum to the maximum (the optical density of the excitation filters decreases accordingly) and then decreases from the maximum to the minimum.

| Excitation Filter 1, Optical Density | Excitation Filter 2, Optical density | Emission Filter, Optical density | Exposure Time (ms) | f (Hz) | Shutter (0 - opened, 1 - closed) |
|---|---|---|---|---|---|
| 4 | 0.3 | 0 | 1000 | 80000000 | 0 |
| 4 | 0 | 0 | 1000 | 80000000 | 0 |
| 3 | 0.3 | 0 | 200 | 80000000 | 0 |
| 3 | 0 | 0 | 200 | 80000000 | 0 |
| 2 | 0.3 | 1 | 200 | 80000000 | 0 |
| 2 | 0 | 1 | 100 | 80000000 | 0 |
| 1 | 0.3 | 2 | 100 | 80000000 | 0 |
| 1 | 0 | 2 | 70 | 80000000 | 0 |
| 0 | 0.3 | 3 | 70 | 80000000 | 0 |
| 0 | 0 | 3 | 70 | 80000000 | 0 |
| 0 | 0.3 | 3 | 70 | 80000000 | 0 |
| 1 | 0 | 2 | 70 | 80000000 | 0 |
| 1 | 0.3 | 2 | 70 | 80000000 | 0 |
| 2 | 0 | 1 | 100 | 80000000 | 0 |
| 2 | 0.3 | 1 | 100 | 80000000 | 0 |
| 3 | 0 | 0 | 200 | 80000000 | 0 |
| 3 | 0.3 | 0 | 200 | 80000000 | 0 |
| 4 | 0 | 0 | 1000 | 80000000 | 0 |
| 4 | 0.3 | 0 | 1000 | 80000000 | 0 |
| 4 | 0 | 0 | 1000 | 80000000 | 0 |

In order to measure many cycles of PL vs I, the measurement table presented above is repeated the required number of cycles. For example, the table shown in **Supplementary Table 2** was repeated 584 times to perform the long hysteresis measurement displayed in **Supplementary Fig.3** and **Fig. 2b** in the main text.

The automated measurement setup allows to realize any excitation sequences (**Supplementary Table 3**). An example of a non-standard excitation sequence is given below:

1) The sample is excited at a high intensity condition (similar to voltage pre-bias in case of memristors);
2) PL vs I is scanned 3 times over a desired range;
3) The sample is left to rest for 1 hour in dark (this is to allow its relaxation towards the initial state), (Note: status 1 in the table means the shutter on the laser is closed and the sample is not exposed to light);

4) PL vs I scanned again 3 times over the desired range.

**Supplementary Table 3.** An example of table for a "non-trivial" PL vs I scan. Initially the sample is exposed to high light intensity for 10 seconds, then PL vs I scan is performed, then the sample rests for 1 hour in dark. Finally, 3 PL vs I scans are again performed.

| Excitation Filter 1, Optical Density | Excitation Filter 2, Optical density | Emission Filter, Optical density | Exposure Time (ms) | f (Hz) | Shutter (0 - opened, 1 - closed) |
|---|---|---|---|---|---|
| 0 | 0 | 4 | 10000 | 80000000 | 0 |
| 4 | 0 | 0 | 1000 | 80000000 | 0 |
| 3 | 0 | 0 | 200 | 80000000 | 0 |
| 2 | 0 | 1 | 100 | 80000000 | 0 |
| 1 | 0 | 2 | 70 | 80000000 | 0 |
| 0 | 0 | 3 | 70 | 80000000 | 0 |
| 1 | 0 | 2 | 70 | 80000000 | 0 |
| 2 | 0 | 1 | 100 | 80000000 | 0 |
| 3 | 0 | 0 | 200 | 80000000 | 0 |
| 4 | 0 | 0 | 1000 | 80000000 | 0 |
| 4 | 0 | 0 | 3600000 | 80000000 | 1 |
| 4 | 0 | 0 | 1000 | 80000000 | 0 |
| 3 | 0 | 0 | 200 | 80000000 | 0 |
| 2 | 0 | 1 | 100 | 80000000 | 0 |
| 1 | 0 | 2 | 70 | 80000000 | 0 |
| 0 | 0 | 3 | 70 | 80000000 | 0 |
| 1 | 0 | 2 | 70 | 80000000 | 0 |
| 2 | 0 | 1 | 100 | 80000000 | 0 |
| 3 | 0 | 0 | 200 | 80000000 | 0 |
| 4 | 0 | 0 | 1000 | 80000000 | 0 |

(First green block repeated x3; second green block repeated x3)

## 2.4 Time-resolved PL under pulsed burst excitation measured by TCSPC.

To record the time-resolved PL of memlumors, a time-correlated single photon counting (TCSPC) setup based on a picosecond diode laser, multichannel picosecond diode laser driver SEPIA 828 (PicoQuant) and time counting device PicoHarp 300 (PicoQuant) was utilized in a non-standard operation regime.

In a traditional TCSPC scheme, a pulsed laser with a fixed repetition rate is used. It means that the sample is excited by one short laser pulse and this process is repeated with the certain repetition rate. For example, if the repetition rate is 100 kHz, the period between each two pulses is T=10 μs. During each period T, a trigger signal generated by the laser driver arrives to the channel 1 (start pulse) of the time counting device. After detecting this signal, the counting device measures the time until the stop pulse arrives to its channel 2 from the photodetector (one detected photon generates an electrical pulse). The histogram of the delays between the start and the stop pulses represents the shape of the PL decay [10].

Contrary to this standard operation regime, the memlumors are excited by a burst of several closely separated laser pulses repeated with a repetition period T. Such pulse time-pattern is created by a Sepia 828 oscillator software. For example, the experiment shown in Fig. 2c was carried out using the following time sequence of the pulses per one period T (see also **Supplementary Fig.2**):

The basic frequency was set to 20 MHz resulting in a 50 ns separation between the pulses. The entire pulse sequence is the following:

- The 1st pulse creates the start signal for the counting electronics;
- The next 40 pulses are sent to the laser driver and produce the burst of 40 laser pulses which excite the sample;
- The next 10 pulses are skipped creating a delay of 500 ns;
- The next single pulse is sent to the laser driver and produces one laser pulse which also excites the sample;
- The next 8000 pulses are skipped to create a large delay.

The entire sequence is repeated over 100 s to accumulate the PL response curve.
The total time of the sequence is 1+40+8000 = 8041 pulses, or 8041*50 ns = 402 μs. This corresponds to the repetition frequency of 1/402 μs = 2 488 Hz.
This excitation pulse sequence was used to measure the PL response presented in **Supplementary Fig.2** and **Fig. 2d** in the main manuscript.

There are several important practical issues concerning the count rates in such an excitation scheme. Older generations of TCSPC electronics operated in a manner that after detection of a stop signal the system was waiting until the next start pulse to start the counting again. As a consequence, no more than just one stop pulse could be detected. Thus, in order to obtain a histogram which would reflect the real PL decay curve without the negative influence of the pile-up effect, one needed to limit the number of stop pulses per start to a value that was much less than one. In practice, the "rule of thumb" was that the detector count rate should be less than 5% of the start pulse rate equal to repetition rate of the laser pulses.

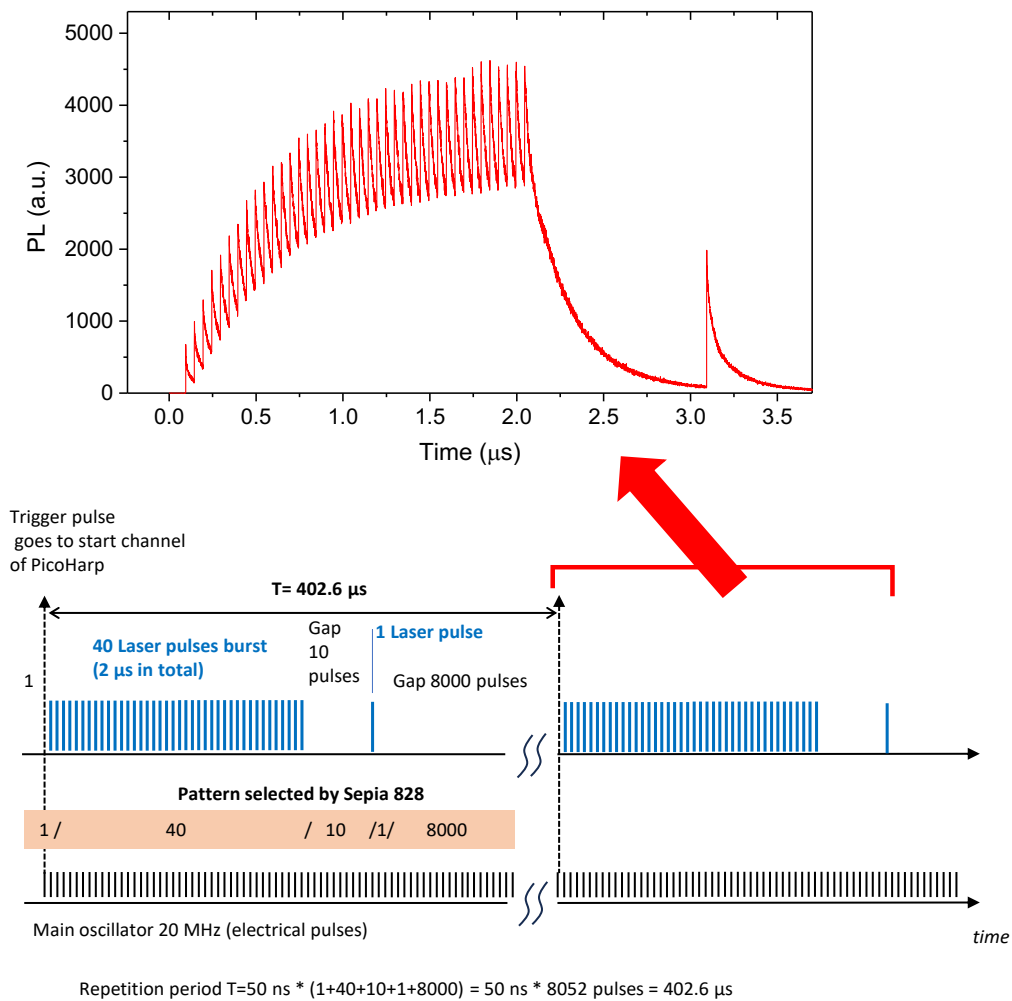

**Supplementary Fig. 2.** TCSPC technique with excitations by a burst of laser pulses.

Following this "rule of thumb" with the burst excitation scheme discussed above, the maximum count rate of the signal should be limited to 5%*2488Hz= 124 counts per second only. Taking into account that the dark counts of the detector (PMA Hybrid detector, PicoQuant) is approximately 200 s$^{-1}$, this signal is below the noise level. Thus, with this limited detection count rate it is not possible to obtain a curve with such a high signal to noise ratio for low excitation intensities, such as the one presented in **Supplementary Fig.2**. However, this "rule of thumb" is not applicable to the experimental conditions applied to the memlumors, and as will be explain in the following, a much higher signal count rates can be obtained.

This is made possible by the fact that modern generations of TCSPC systems are able to count the time difference relatively to the start pulse for many detection events per repetition period as long as the time gap to the next photon is substantially larger than the detection dead time. So, the pile-up effect in its classical sense is only present when the repetition period of the laser pulses or the fluorescence decay itself is shorter than the deadtime. For example, for our system (PicoHarp 300) the deadtime is approximately 100 ns. Based on this, only for laser repetition rates larger than 1 MHz (time gap to the next pulse is <100 ns) does the 5% rule apply.

In the measurement scenario applied to the memlumors, the repetition rate was as low as 2.5KHz (400 μs repetition period). At the same time, the PL signal is spread over the time window of approximately 3 μs. Both these times are substantially larger than the detection deadtime (100 ns). Therefore, several photons can be detected per repetition period without any pile-up effect. This leads to a count range of several kHz which is much larger than the noise level. The maximum safe count rate depends on the particular shape of the PL response curve, and a detailed discussion of this topic is beyond the scope of this study. In practice, test experiments were performed by increasing the count rate to identify the point at which distortions of the PL response due to the pile up effect appeared, and then a rate of at least 5 times smaller was used for sample characterisation.

## 2.5 PLQY(f,P) mapping

For the PLQY(f,P) mapping experiments [8,11] the sample is excited by a pulsed laser at repetition rate $f$ [s$^{-1}$] and pulse fluence P$_i$ [photons/cm$^2$] which are controlled by the laser driver and a neutral density filter wheel, respectively. The sample PL intensity is measured for each combination of $f$ and P. We used a CCD camera to measure the PL integrated over 30 μm laser excitation spot. Because the entire system is calibrated, the PL intensity can be converted to an external PLQY [1,2]. To acquire the PLQY($f$,P) map, PL is measured for a laser excitation spanning over 4 orders of magnitude (from *ca.* 10$^8$ to 10$^{12}$ photons/cm$^2$/pulse) in 4 steps with power fluences P$_i$ - P1, P2, P3, P4 and P5 (each step changes the fluence approximately 10 times, see **Supplementary Table 1**) and almost 7 orders of magnitude in pulse repetition rate, i.e., from 10 Hz to 80 MHz (**Supplementary Table 4, Supplementary Table 5)**. For each pulse fluence P, the repetition rate, $f$, is scanned across the entire range. Usually, a complete map consists of 50-150 data points.

All the data points of the PLQY($f$, P) map are acquired automatically because the setup is fully controlled by home-developed LabVIEW program. It executes the experiment according to a pre-loaded table of parameters for the data acquisition ($f$, P, shutter timing, acquisition time of the camera, filters, *etc.*). To minimize the sample exposure, the shutter is synchronized with camera acquisition to allow the laser beam to irradiate the sample during the PL acquisition only. The program automatically saves the PL images, which are later processed using another program to yield the PLQY($f$,P) map. This ensures that the data acquisition conditions are fully reproducible to be able to repeat exactly the same experiment with another sample. This is of curial importance for light-sensitive materials like metal-halide perovskites and memlumors in general. The complete measurement takes from 1 to 3 hours, where the longest time is required to acquire data for low P and low $f$ values since exposure times as long as several minutes per data point are often essential.

# Supplementary Note 3.
# Measurements of the Long-Term Memory Effects in Memlumors

### 3.1 PL vs intensity hysteresis plot for CsPbBr$_3$ film memlumor

The memlumor photoluminescence (PL) vs the excitation intensity (I in W/cm$^2$) hysteresis curves can be measured using the methodology described in **S2.3.** The table for the experiment shown in **Fig. 2b** in the main text consists of the **Supplementary Table 2** (which describes one cycle) which is repeated 584 times. During the scan, the evolution of the CsPbBr$_3$ polycrystalline film sample with the cycle number is recorded. Note that going through quite high excitation light power density (160 W/cm$^2$, f=80MHz, P5) was intentionally chosen to induce long-term photochemical processes in the sample in each cycle. During the first scans the PL(I) hysteresis shifts down, representing a decrease in the relative PLQY that is defined as PL/I (**Supplementary Fig.3a**). After approximately 20 cycles, the hysteresis loop stabilizes and stays stable over the next 200 cycles (**Supplementary Fig.3a**). After that, the hysteresis loop starts to drift again (increasing of PLQY) until the end of the experiment (364 cycles, **Supplementary Fig.3a**).

Due to the linear scale, it is difficult to observe that PL under low excitation power also evolves during the cycles. It is illustrated by plotting the evolution of PL excited by a 1.4 W/cm$^2$ power density (**Supplementary Fig.3b**). This evolution is complex (highly non-linear) and reflects the history of the sample excitation during each repeated scan of excitation conditions (**Supplementary Table 2**).
The evolution of memlumor's PL can also be visualized by plotting the PLQY at the minimum 0.005 W/cm$^2$ and maximum 160 W/cm$^2$ excitation points during the cycling (**Supplementary Fig.3e**). The stabilization of the PLQY (**Supplementary Fig.3e**) occurs for both excitation conditions at the same time (horizontal regions, marked in the figure). This stable region corresponds to the stable 200 cycles of the hysteresis loop displayed in **Supplementary Fig.3c,d.**

The presence of hysteresis in memlumors is representative of volatile memory effects. When excited by the high excitation powers, the state $\vec{S}$ of the sample is changed, while during the low excitation intensity region, the sample recovers its previous state. Correspondingly, remaining at low light excitation conditions has the effect of memory erasing for the CsPbBr$_3$ memlumors.

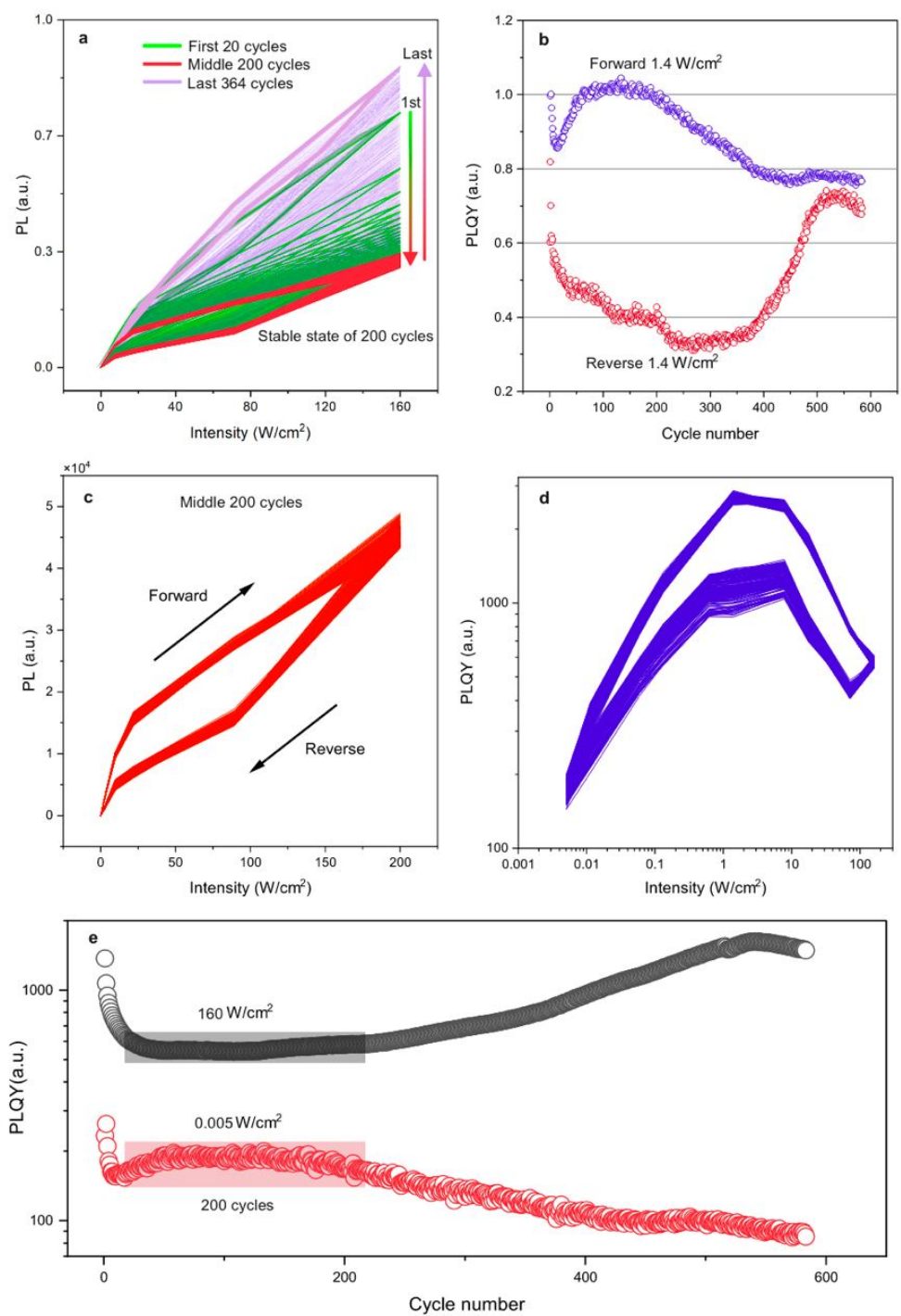

**Supplementary Fig. 3.** Long-term memory measurements of CsPbBr$_3$ film memlumor. a) The full hysteresis plot consisting of 584 repetitive cycles of the excitation from 0.005 to 160 W/cm$^2$ (80 MHz repetition rate). b) The evolution of PLQY at 1.4 W/cm$^2$, forward scan and reverse scan are separated. c) PL versus Intensity plot for the stable 200 cycles of total experiment with the corresponding plot of relative PLQY shown in d). e) The evolution of relative PLQY at the maximum and the minimum intensity of the PL vs I scan. The region of stabilization (200 cycles, marked) is clearly visible.

## 3.2 Long-term volatile memory of the CsPbBr₃ film memlumor

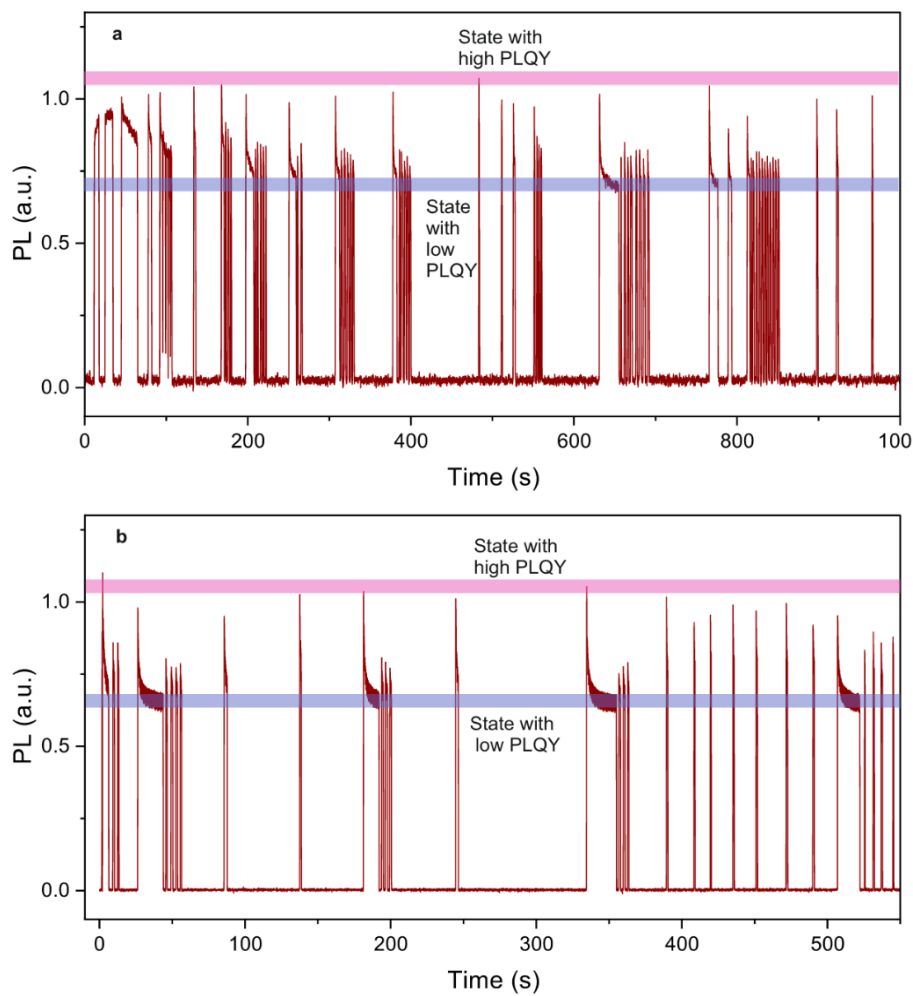

**Supplementary Fig. 4.** Time traces of PL evolution under interrupted excitation of the CsPbBr₃ film. The sample was excited by bursts of 1000 pulses (pulse repetition rate 20 MHz, 220 nJ/cm² per pulse, total burst length = 50 μs) repeated each 21000*50ns=1.05 ms (1000/20 000 = 5% duty cycle). The initially high PL (the state with high PLQY) decreases under constant excitation to a state with a low PLQY. If we allow the memlumor to "rest", the sample PL returns to its initial value. However, when the resting time is not enough, we can probe the state of low PLQY again. This volatile reversible change of the memlumor's state happens at the time scale of tens of seconds.

## 3.3 Long-term memory of the MAPbI$_3$ film memlumor

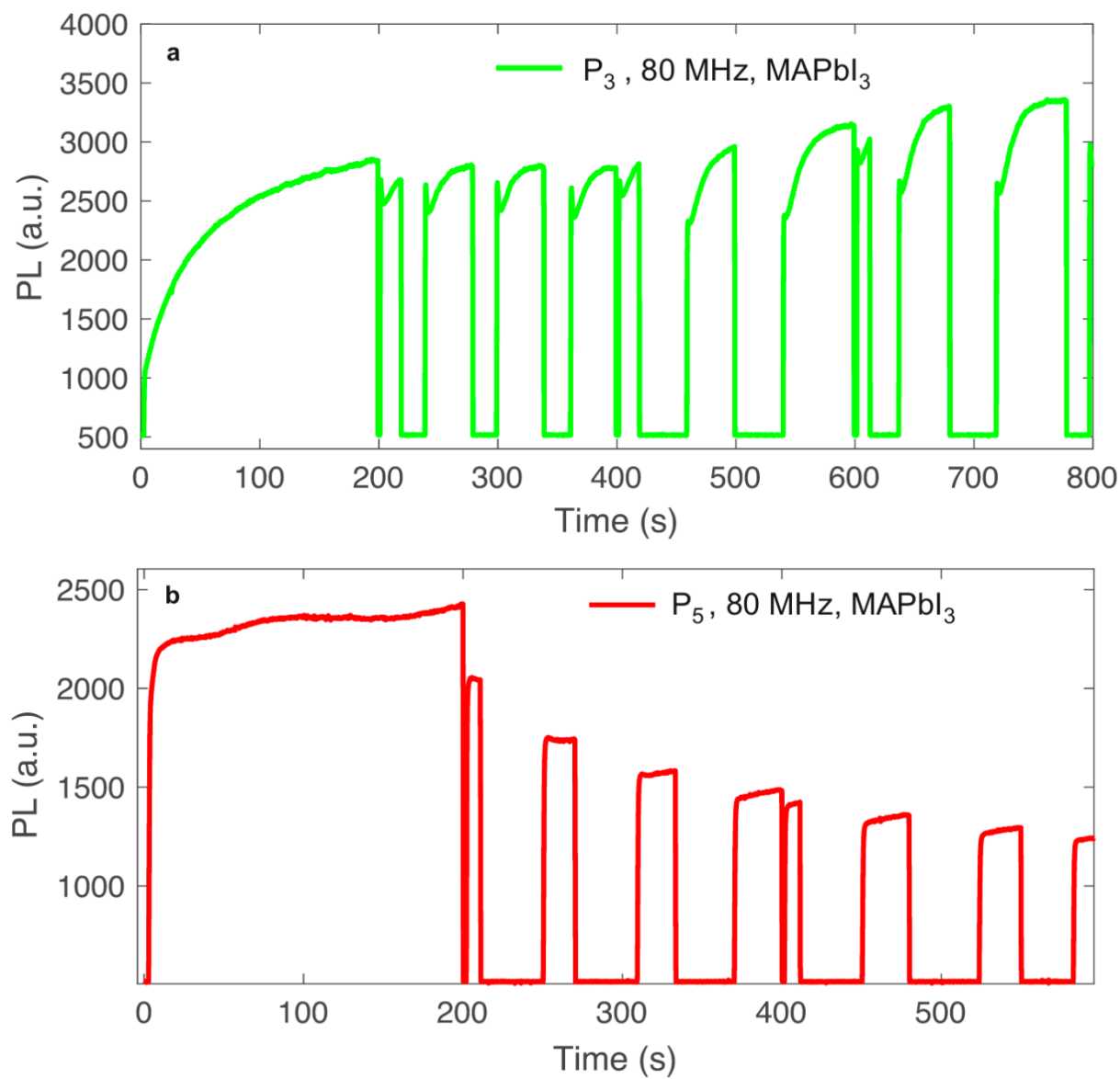

**Supplementary Fig. 5.** Time traces of the PL evolution under interrupted excitation for the MAPbI$_3$ thin film. The excitation conditions: 80 MHz pulse repetition rate with pulse fluences P$_3$ for a) and P$_5$ for b) fluence. The volatile behavior of the MAPbI$_3$ memlumor is visible for the P$_3$ fluence. For the case of the P$_5$ fluence (b), the behavior is more complicated.

## 3.4 Control of the long-term memory by varying the light input pulse sequence

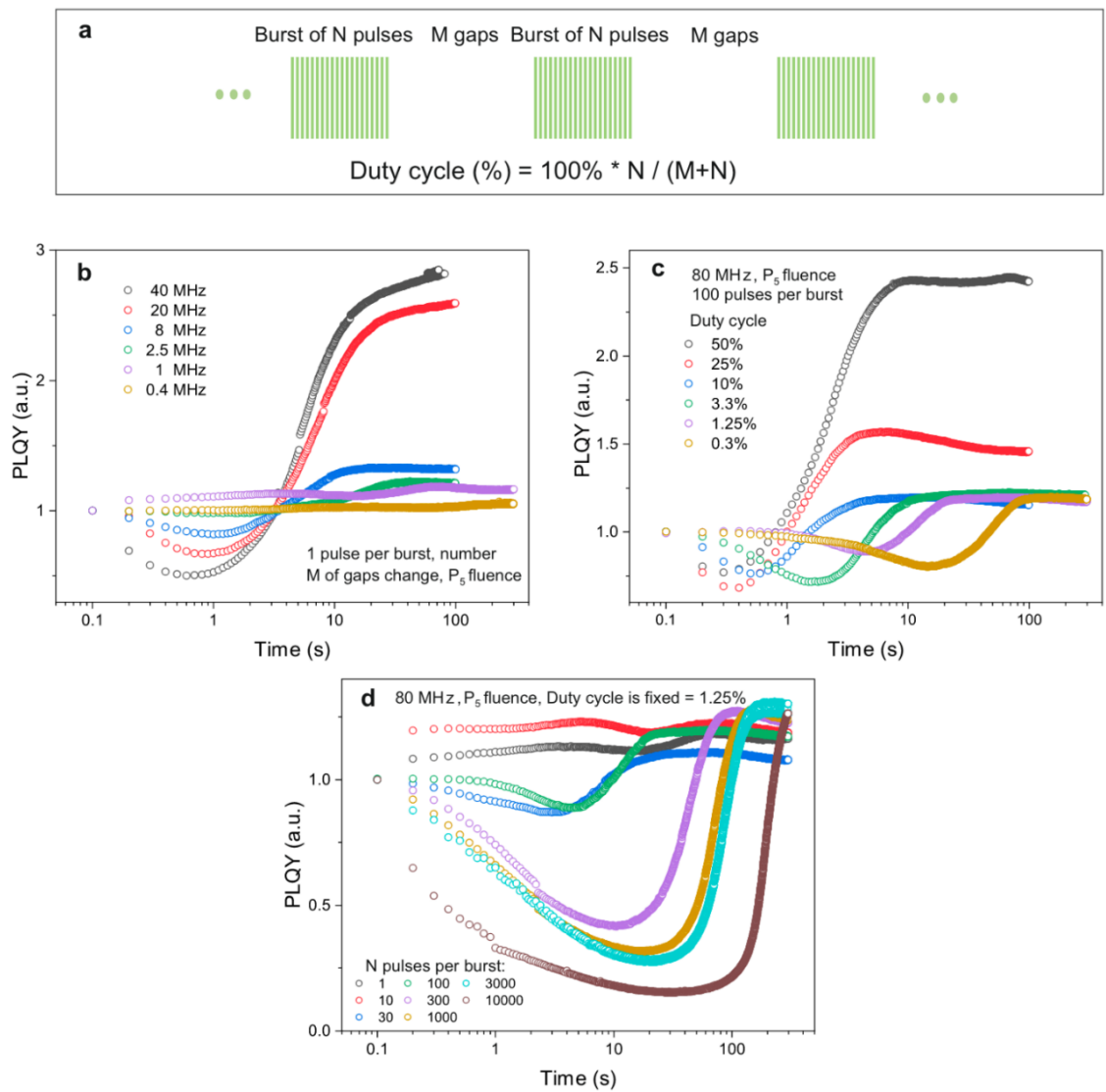

**Supplementary Fig. 6.** Control of the long-term memory of the CsPbBr$_3$ film memlumor by exciting with pulse fluence P$_5$ in a pulse burst regime. a) Schematics of the experiment with bursts. The duty cycle of a burst depends on the ratio between the number of pulses per burst N and the sum of the number of the gaps M and number of the pulses per burst N. The duty cycle is a tool for controlling of the long-term memory. b) For one pulse per burst and varied number of gaps, duty cycle represents the pulse repetition rate. Excitation with high repetition rate initiates changes in memlumor´s relative PLQY. c) At constant frequency of 80 MHz and 100 pulses per burst, the number of gaps changes the duty cycle. The smaller the duty cycle, the slower the process of the relative PLQY change. d) The duty cycle is fixed meaning that the average excitation intensity is constant, but the number of pulses per burst changes. The shape of the relative PLQY versus irradiation time dependence is significantly dependent on the duty cycle despite the average excitation intensity being the same.

# Supplementary Note 4. Description of the PLQY (f, P) Mapping for the CsPbBr₃ Film Memlumor.

The goal of the PLQY(f,P) mapping is to obtain the parameters of the defect states $\vec{S} = (k_t, k_n, N_t)$ and all other rate constants of the SRH+ model by probing the PL upon a change in the excitation conditions. This changes the components of the state vector responsible for the short-term memory $(n, n_t)$ over a very broad range. Note that because these components react to any change of the excitation conditions very fast (much faster than a microsecond), the PL is always measured at the quasi-steady state conditions for $(n, n_t)$ (see **Supplementary Note 7.3** for a detailed discussion regarding the time *t'*) under the standard experimental conditions [8] when the exposure time per data point (a combination of *f* and *P*) is larger than 1 ms. So, PLQY(f,P) mapping makes it possible to obtain the components of the state vector $\vec{S} = (k_t, k_n, N_t)$ assuming that these components are not affected by the experimental procedure, which mathematically means that:

$$\frac{d\vec{S(f,P)}}{dt} = 0 \qquad \text{S4.1}$$

However, this assumption contradicts the knowledge of CsPbBr₃ as a photosensitive material, where the parameters of the defect states are light sensitive. In other words, it is known that the experiment itself may influence the result of the measurement (so-called observer effect, see details in reference [11]). Considering that the PLQY(f,P) measurements involve both very low and very high excitation power densities, extra care should be taken so that the results are not affected by a constant evolution of the material under light irradiation (excitation history). This means it is important to consider that the same data point measured with different irradiation time (exposure time) might result in a different value of PLQY.

**Supplementary Fig.7a** shows the PLQY(f,P) map obtained for the CsPbBr₃ film studied here using the standard experimental protocol used previously for these type of measurements.[8] The parameters used to measure each experimental point are shown in **Supplementary Table 4**. A simple visual inspection of the PLQY(f,P) plot reveals the presence of significant sample instability.[11] This is evidenced by the appearance of a non-constant PLQY where the single pulse excitation regime is expected and the absence of common quasi-CW regime for all the pulse fluences (see **Supplementary Note** 7 and references [8,11] as well as the labels in **Supplementary Fig.7a**).

To understand this, it is helpful to roughly split the entire PLQY(f,P) map into two regions: **Region 1**, in which the averaged excitation power density is mild or low and **Region 2**, in which the averaged power density is high. **Supplementary Table 4** lists the experimental parameters for the measurements. The vector $\vec{S} = (k_t, k_n, N_t)$ is likely to change in Region 2, however the impact of this change is more clearly observed in Region 1, mostly like due to the fact that non-radiative recombination is dominant in this region. Hence, to track the change in the state of the memlumor during the measurements, a reference point is measured at a fluence $P_2$ at 80 MHz (inside Region 1) after each point of the PLQY (f,P) map.

The PL intensity at this reference points should be constant if the sample is stable, however, this is clearly not the case as can be clearly seen in **Supplementary Fig.7b**. The first and the last value of the reference point varies by two orders of magnitude. The evolution of the reference point (**Supplementary Fig.7b**) shows that entering Region 2 influences the sample significantly. Hence the values of the $\vec{S}$ vector for Region 1 and Region 2 must be very different, and as a result, the PLQY(f,P) map is affected by the sample scanning history and cannot be fitted correctly because equation S4.1 does not hold ($\frac{d\overrightarrow{S(f,P)}}{dt} \neq 0$). Indeed, the shape of the PLQY (f,P) map contains features that are not compatible with any charge recombination model with constant parameters (see further explanation in **Supplementary Fig.7a**).

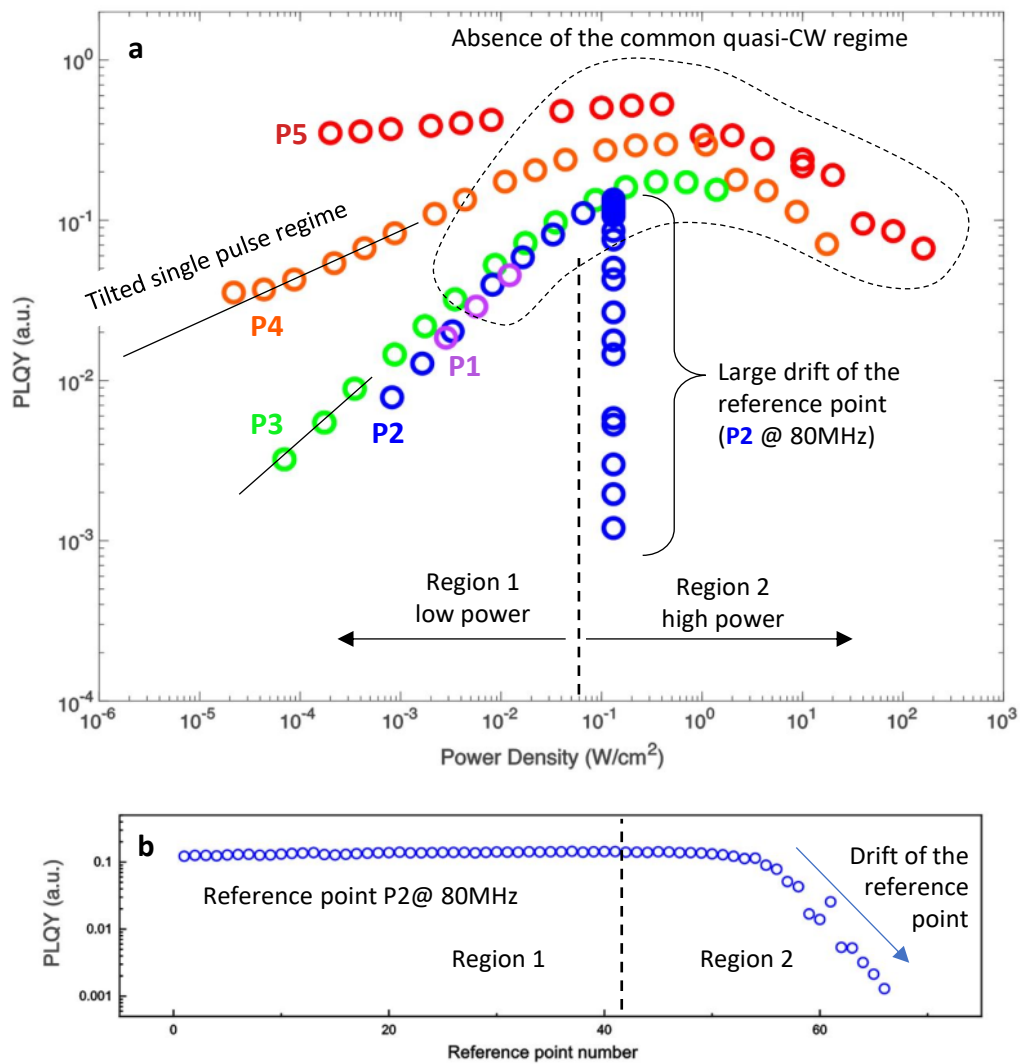

**Supplementary Fig. 7**. a) Standard [8] measurement of the PLQY(f,P) map for CsPbBr$_3$ polycrystalline film. Two regions are separated: Region 1, where the input power density is low, and Region 2 where the input power density is high (> 0.05 W/cm$^2$). The map exhibits very strong artifacts due to sample instability (see the notes). b) Evolution of the reference point.

**Supplementary Table 4.** The conditions for acquiring of the experimental points listed in measurement order (from 1 to 125) in the experiment shown in **Supplementary Fig.7**. The two regions of low and high intensity are separated. The rows showing the reference points are filled with the white color to separate them from the rows (filled by light green) showing the actual points of the PLQY(f, P) map. The font color highlights the fluences P1-P5 according to the color scheme used in **Supplementary Fig.7a.**

### Region 1

| Point number | Excitation Filter 1, Optical Density | Excitation Filter 2, Optical density | Emission Filter, Optical density | Exposure Time (ms) | f (Hz) | Shutter (0 - opened, 1 - closed) |
|---|---|---|---|---|---|---|
| 1 | 3 | 0 | 0 | 100 | 80000000 | 0 |
| 2 | 4 | 0 | 0 | 30000 | 20000000 | 0 |
| 3 | 3 | 0 | 0 | 100 | 80000000 | 0 |
| 4 | 4 | 0 | 0 | 15000 | 40000000 | 0 |
| 5 | 3 | 0 | 0 | 100 | 80000000 | 0 |
| 6 | 4 | 0 | 0 | 8000 | 80000000 | 0 |
| 7 | 3 | 0 | 0 | 100 | 80000000 | 0 |
| 8 | 3 | 0 | 0 | 30000 | 500000 | 0 |
| 9 | 3 | 0 | 0 | 100 | 80000000 | 0 |
| 10 | 3 | 0 | 0 | 20000 | 1000000 | 0 |
| 11 | 3 | 0 | 0 | 100 | 80000000 | 0 |
| 12 | 3 | 0 | 0 | 10000 | 2000000 | 0 |
| 13 | 3 | 0 | 0 | 100 | 80000000 | 0 |
| 14 | 2 | 0 | 0 | 100000 | 4000 | 0 |
| 15 | 3 | 0 | 0 | 100 | 80000000 | 0 |
| 16 | 2 | 0 | 0 | 50000 | 10000 | 0 |
| 17 | 3 | 0 | 0 | 100 | 80000000 | 0 |
| 18 | 2 | 0 | 0 | 20000 | 20000 | 0 |
| 19 | 3 | 0 | 0 | 100 | 80000000 | 0 |
| 20 | 2 | 0 | 0 | 20000 | 50000 | 0 |
| 21 | 3 | 0 | 0 | 100 | 80000000 | 0 |
| 22 | 2 | 0 | 0 | 10000 | 100000 | 0 |
| 23 | 3 | 0 | 0 | 100 | 80000000 | 0 |
| 24 | 2 | 0 | 0 | 5000 | 200000 | 0 |
| 25 | 3 | 0 | 0 | 100 | 80000000 | 0 |
| 26 | 1 | 0 | 0 | 100000 | 100 | 0 |
| 27 | 3 | 0 | 0 | 100 | 80000000 | 0 |
| 28 | 1 | 0 | 0 | 100000 | 200 | 0 |
| 29 | 3 | 0 | 0 | 100 | 80000000 | 0 |
| 30 | 1 | 0 | 0 | 50000 | 400 | 0 |
| 31 | 3 | 0 | 0 | 100 | 80000000 | 0 |
| 32 | 1 | 0 | 0 | 20000 | 1000 | 0 |
| 33 | 3 | 0 | 0 | 100 | 80000000 | 0 |
| 34 | 1 | 0 | 0 | 10000 | 2000 | 0 |
| 35 | 3 | 0 | 0 | 100 | 80000000 | 0 |
| 36 | 1 | 0 | 0 | 5000 | 4000 | 0 |
| 37 | 3 | 0 | 0 | 100 | 80000000 | 0 |
| 38 | 1 | 0 | 0 | 2000 | 10000 | 0 |
| 39 | 3 | 0 | 0 | 100 | 80000000 | 0 |
| 40 | 1 | 0 | 0 | 1000 | 20000 | 0 |
| 41 | 3 | 0 | 0 | 100 | 80000000 | 0 |
| 42 | 0 | 0 | 0 | 5000 | 100 | 0 |
| 43 | 3 | 0 | 0 | 100 | 80000000 | 0 |
| 44 | 0 | 0 | 0 | 5000 | 200 | 0 |
| 45 | 3 | 0 | 0 | 100 | 80000000 | 0 |
| 46 | 0 | 0 | 0 | 2000 | 400 | 0 |
| 47 | 3 | 0 | 0 | 100 | 80000000 | 0 |
| 48 | 0 | 0 | 0 | 1000 | 1000 | 0 |
| 49 | 3 | 0 | 0 | 100 | 80000000 | 0 |
| 50 | 0 | 0 | 0 | 500 | 2000 | 0 |
| 51 | 3 | 0 | 0 | 100 | 80000000 | 0 |
| 52 | 0 | 0 | 0 | 300 | 4000 | 0 |
| 53 | 3 | 0 | 0 | 100 | 80000000 | 0 |
| 54 | 3 | 0 | 0 | 5000 | 5000000 | 0 |
| 55 | 3 | 0 | 0 | 100 | 80000000 | 0 |
| 56 | 2 | 0 | 0 | 2000 | 500000 | 0 |
| 57 | 3 | 0 | 0 | 100 | 80000000 | 0 |
| 58 | 1 | 0 | 0 | 1000 | 50000 | 0 |
| 59 | 3 | 0 | 0 | 100 | 80000000 | 0 |
| 60 | 3 | 0 | 0 | 2000 | 10000000 | 0 |
| 61 | 3 | 0 | 0 | 100 | 80000000 | 0 |
| 62 | 2 | 0 | 0 | 1000 | 1000000 | 0 |
| 63 | 3 | 0 | 0 | 100 | 80000000 | 0 |
| 64 | 1 | 0 | 0 | 500 | 100000 | 0 |
| 65 | 3 | 0 | 0 | 100 | 80000000 | 0 |
| 66 | 0 | 0 | 0 | 70 | 10000 | 0 |
| 67 | 3 | 0 | 0 | 100 | 80000000 | 0 |
| 67 | 3 | 0 | 0 | 1000 | 20000000 | 0 |
| 68 | 3 | 0 | 0 | 100 | 80000000 | 0 |
| 68 | 3 | 0 | 0 | 500 | 40000000 | 0 |
| 69 | 3 | 0 | 0 | 100 | 80000000 | 0 |
| 70 | 2 | 0 | 0 | 500 | 2000000 | 0 |
| 71 | 3 | 0 | 0 | 100 | 80000000 | 0 |
| 72 | 1 | 0 | 0 | 500 | 200000 | 0 |
| 73 | 3 | 0 | 0 | 100 | 80000000 | 0 |
| 74 | 0 | 0 | 0 | 70 | 20000 | 0 |

### Region 2

| Point number | Excitation Filter 1, Optical Density | Excitation Filter 2, Optical density | Emission Filter, Optical density | Exposure Time (ms) | f (Hz) | Shutter (0 - opened, 1 - closed) |
|---|---|---|---|---|---|---|
| 75 | 3 | 0 | 0 | 100 | 80000000 | 0 |
| 76 | 2 | 0 | 0 | 500 | 5000000 | 0 |
| 77 | 3 | 0 | 0 | 100 | 80000000 | 0 |
| 78 | 1 | 0 | 0 | 200 | 500000 | 0 |
| 79 | 3 | 0 | 0 | 100 | 80000000 | 0 |
| 80 | 0 | 0 | 1 | 70 | 50000 | 0 |
| 81 | 3 | 0 | 0 | 100 | 80000000 | 0 |
| 82 | 3 | 0 | 0 | 100 | 80000000 | 0 |
| 83 | 3 | 0 | 0 | 100 | 80000000 | 0 |
| 84 | 2 | 0 | 0 | 200 | 10000000 | 0 |
| 85 | 3 | 0 | 0 | 100 | 80000000 | 0 |
| 86 | 1 | 0 | 0 | 100 | 1000000 | 0 |
| 87 | 3 | 0 | 0 | 100 | 80000000 | 0 |
| 88 | 0 | 0 | 1 | 70 | 100000 | 0 |
| 89 | 3 | 0 | 0 | 100 | 80000000 | 0 |
| 90 | 2 | 0 | 0 | 100 | 20000000 | 0 |
| 91 | 3 | 0 | 0 | 100 | 80000000 | 0 |
| 92 | 1 | 0 | 1 | 200 | 2000000 | 0 |
| 93 | 3 | 0 | 0 | 100 | 80000000 | 0 |
| 94 | 0 | 0 | 1 | 70 | 200000 | 0 |
| 95 | 3 | 0 | 0 | 100 | 80000000 | 0 |
| 96 | 2 | 0 | 1 | 200 | 40000000 | 0 |
| 97 | 3 | 0 | 0 | 100 | 80000000 | 0 |
| 98 | 1 | 0 | 1 | 100 | 5000000 | 0 |
| 99 | 3 | 0 | 0 | 100 | 80000000 | 0 |
| 100 | 0 | 0 | 2 | 100 | 500000 | 0 |
| 101 | 3 | 0 | 0 | 100 | 80000000 | 0 |
| 102 | 2 | 0 | 1 | 100 | 80000000 | 0 |
| 103 | 3 | 0 | 0 | 100 | 80000000 | 0 |
| 104 | 1 | 0 | 2 | 500 | 10000000 | 0 |
| 105 | 3 | 0 | 0 | 100 | 80000000 | 0 |
| 106 | 0 | 0 | 2 | 70 | 1000000 | 0 |
| 107 | 3 | 0 | 0 | 100 | 80000000 | 0 |
| 108 | 1 | 0 | 2 | 200 | 20000000 | 0 |
| 109 | 3 | 0 | 0 | 100 | 80000000 | 0 |
| 110 | 0 | 0 | 3 | 200 | 2000000 | 0 |
| 111 | 3 | 0 | 0 | 100 | 80000000 | 0 |
| 112 | 1 | 0 | 2 | 200 | 40000000 | 0 |
| 113 | 3 | 0 | 0 | 100 | 80000000 | 0 |
| 114 | 0 | 0 | 3 | 100 | 5000000 | 0 |
| 115 | 3 | 0 | 0 | 100 | 80000000 | 0 |
| 116 | 1 | 0 | 2 | 100 | 80000000 | 0 |
| 117 | 3 | 0 | 0 | 100 | 80000000 | 0 |
| 118 | 0 | 0 | 3 | 100 | 10000000 | 0 |
| 119 | 3 | 0 | 0 | 100 | 80000000 | 0 |
| 120 | 0 | 0 | 4 | 200 | 20000000 | 0 |
| 121 | 3 | 0 | 0 | 100 | 80000000 | 0 |
| 122 | 0 | 0 | 4 | 200 | 40000000 | 0 |
| 123 | 3 | 0 | 0 | 100 | 80000000 | 0 |
| 124 | 0 | 0 | 4 | 70 | 80000000 | 0 |
| 125 | 3 | 0 | 0 | 100 | 80000000 | 0 |

**Supplementary Table 5.** The experimental points located in the order of their acquisition (from 1 to 42) in the experiment shown in **Supplementary Fig. S8a, b**. The two regions of low and high intensity are separated. The rows showing the reference points are filled with white color to separate the from the rows (filled with light green) showing the actual points of the PLQY(f, P) map. The font color highlights the fluences P1-P5 according to the standard color scheme used in **Supplementary Fig. S8a, b.**

| Point number | Excitation Filter 1, Optical Density | Excitation Filter 2, Optical density | Emission Filter, Optical density | Exposure Time (ms) | f (Hz) | Shutter (0 - opened, 1 - closed) |
|---|---|---|---|---|---|---|
| 1 | 4 | 0 | 0 | 40000 | 1000000 | 0 |
| 2 | 4 | 0 | 0 | 8000 | 5000000 | 0 |
| 3 | 4 | 0 | 0 | 2000 | 10000000 | 0 |
| 4 | 4 | 0 | 0 | 1000 | 20000000 | 0 |
| 5 | 4 | 0 | 0 | 500 | 40000000 | 0 |
| 6 | 4 | 0 | 0 | 250 | 80000000 | 0 |
| 7 | 3 | 0 | 0 | 180000 | 10000 | 0 |
| 8 | 3 | 0 | 0 | 60000 | 30000 | 0 |
| 9 | 3 | 0 | 0 | 20000 | 100000 | 0 |
| 10 | 3 | 0 | 0 | 4000 | 500000 | 0 |
| 11 | 3 | 0 | 0 | 1000 | 2000000 | 0 |
| 12 | 3 | 0 | 0 | 250 | 5000000 | 0 |
| 13 | 3 | 0 | 0 | 140 | 10000000 | 0 |
| 14 | 3 | 0 | 0 | 70 | 20000000 | 0 |
| 15 | 3 | 0 | 0 | 70 | 40000000 | 0 |
| 16 | 3 | 0 | 0 | 70 | 80000000 | 0 |
| 17 | 2 | 0 | 0 | 540000 | 300 | 0 |
| 18 | 2 | 0 | 0 | 180000 | 1000 | 0 |
| 19 | 2 | 0 | 0 | 60000 | 3000 | 0 |
| 20 | 2 | 0 | 0 | 15000 | 10000 | 0 |
| 21 | 2 | 0 | 0 | 4000 | 30000 | 0 |
| 22 | 2 | 0 | 0 | 2000 | 100000 | 0 |
| 23 | 1 | 0 | 0 | 240000 | 10 | 0 |
| 24 | 1 | 0 | 0 | 80000 | 30 | 0 |
| 25 | 1 | 0 | 0 | 40000 | 100 | 0 |
| 26 | 1 | 0 | 0 | 20000 | 300 | 0 |
| 27 | 1 | 0 | 0 | 5000 | 1000 | 0 |
| 28 | 1 | 0 | 0 | 2000 | 3000 | 0 |
| 29 | 1 | 0 | 0 | 1000 | 10000 | 0 |
| 30 | 0 | 0 | 0 | 5000 | 100 | 0 |
| 31 | 0 | 0 | 0 | 3000 | 300 | 0 |
| 32 | 0 | 0 | 0 | 1000 | 1000 | 0 |
| 33 | 0 | 0 | 0 | 300 | 3000 | 0 |
| 34 | 0 | 0 | 0 | 100 | 10000 | 0 |
| 35 | 2 | 0 | 0 | 31,25 | 40000000 | 0 |
| 36 | 2 | 0 | 0 | 15,63 | 80000000 | 0 |
| 37 | 1 | 0 | 0 | 2,50 | 40000000 | 0 |
| 38 | 1 | 0 | 0 | 1,25 | 80000000 | 0 |
| 39 | 0 | 0 | 0 | 0,50 | 20000000 | 0 |
| 40 | 0 | 0 | 0 | 0,25 | 40000000 | 0 |
| 41 | 0 | 0 | 0 | 0,13 | 80000000 | 0 |
| 42 | 4 | 0 | 0 | 250 | 80000000 | 0 |

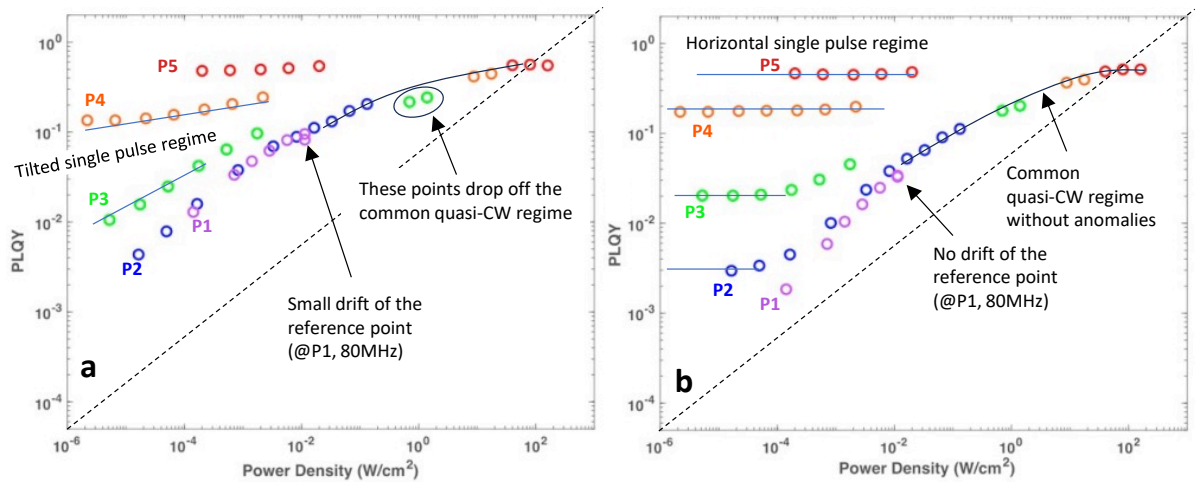

**Supplementary Fig. 8.** The PLQY(f,P) plots with the optimized exposure time according to S**upplementary Table 5.** a) The PLQY(f,P) plot for the initial state of the sample (as-prepared), there are some features revealing sample instability (see notes in the plot). b) PLQY(f,P) map for the sample pre-exposed for 20 seconds at P5 fluence at 80 MHz frequency. By this procedure we prepared $\vec{S_1}$ state of the sample, which was stable enough to measure a reliable PLQY(f,P) map which was used for modelling, see notes oт the figures for details.

Due to technical limitations of the mechanical shutter used in the setup, the minimum exposure time in the standard measurement protocol was 70 ms, which was obviously too long, leading to changes of the sample state when excited at high power densities (Region 2**).** Importantly, from the point of view of the signal strength, a much shorter exposure time would be sufficient to record the PL signal at the high-power density region, for example, 70 ms could be replaced by 1 ms if the shutter would allow it.

Therefore, in order to improve the sample state stability (i.e. to reach a condition which is as close as possible to $\frac{d\overrightarrow{S(f,P)}}{dt} = 0$), the dose of light irradiating the sample was minimized by changing the manner in which the exposure time was set. Instead of using a mechanical shutter, the laser itself was used to determine the irradiation time of the sample. The laser was run in a pulse-burst mode with a very large time interval (1 s) between the bursts (enabled by using a Sepia 828 PicoQuant controller). The mechanical shutter was then synchronized with the laser to open in such a way that within its open time only one pulse burst was able to reach the sample per one measured data point. This approach made it possible to set the exposure for all the data points to the minimal value determined by the signal to noise ratio. For example, the exposure time of the sample to the highest power density 160W/cm$^2$ (P5, 80MHz,) was decreased more than a factor of 500 (from 70 to 0.13 ms). This greatly increased the stability of the sample. However, we note that measurements at some combinations of f and P belonging to Region 2 were avoided since even the minimal possible exposure led to changes in the sample. The protocol for this experiment is provided in **Supplementary Table 5**.

A PLQY (f,P) map measured according to **Supplementary Table 5** is shown in **Supplementary Fig.8a.** The. difference between the 1$^{st}$ and last values of the measured reference point (P1, 80MHz**)** decreases significantly. Still, the sample demonstrated some features incompatible with SRH+ model (see notes on the **Supplementary Fig.8a**). We conclude that the initial, as-prepared state of the sample $\vec{S_0}$, is extremely sensitive to light and is prone to easily change into another state $\vec{S_i}$ during the measurements even when taking significant precautions.

It was found that by exposing the sample to the 160W/cm$^2$ for 20s, it was possible to induce a state (designated as $\vec{S_1}$) which is much more stable as the initial state of the sample. Several experiments shown in the main text (**Fig. 3e, 5c**) were carried out for samples prepared in this way. The PLQY(f,P) plot corresponding to this state is shown in **Supplementary Fig.8b.** In such a state, the reference point was stable and the PLQY(f,P) map did not exhibit any features that could not be well described by the SRH+ model. This data was utilized for modelling using the SRH+ model in order to extract the model parameters (see **Supplementary Note 7.6** and **Supplementary Fig 14 and 15**). Note, however, that the parameters extracted from this data cannot be completely assigned to the $\vec{S_1}$ state. The PLQY(f,P) mapping lasts about 1 h and some relaxation processes of the sample state during this time cannot be excluded.

# Supplementary Note 5. CsPbBr$_3$ Film Memlumor State Vector $\vec{X}$ Evolution from ns to minutes.

The evolution of the PL (or PLQY) driven by the changes in the state vector $\vec{X}$ can be observed for any fluence. In general, it includes both changes of the fast component (n,n$_t$) and the slow component $\vec{S}$ = (k$_n$, k$_t$, N$_t$) of the state vector $\vec{X}$.

**Supplementary Fig.9** shows the time evolution for the state vector $\vec{X}$ at different pulse fluences 17.6 nJ/cm$^2$, 220 nJ/cm$^2$ and for 2000 nJ/cm$^2$ (P3, P4 and P5, respectively). The leak-and-integrate behavior of the CsPbBr$_3$ film memlumor (left part of the **Supplementary Fig.9**) was studied by applying a 20-pulse burst at 80 MHz repetition frequency with the gap of 62.5 µs using the TCSPS setup for recoding the evolution on the nanoseconds time scale and the CCD camera for recording at times longer than 100 ms. These two data sets are scaled to the same units using a calibration procedure performed in between the two detection modes.

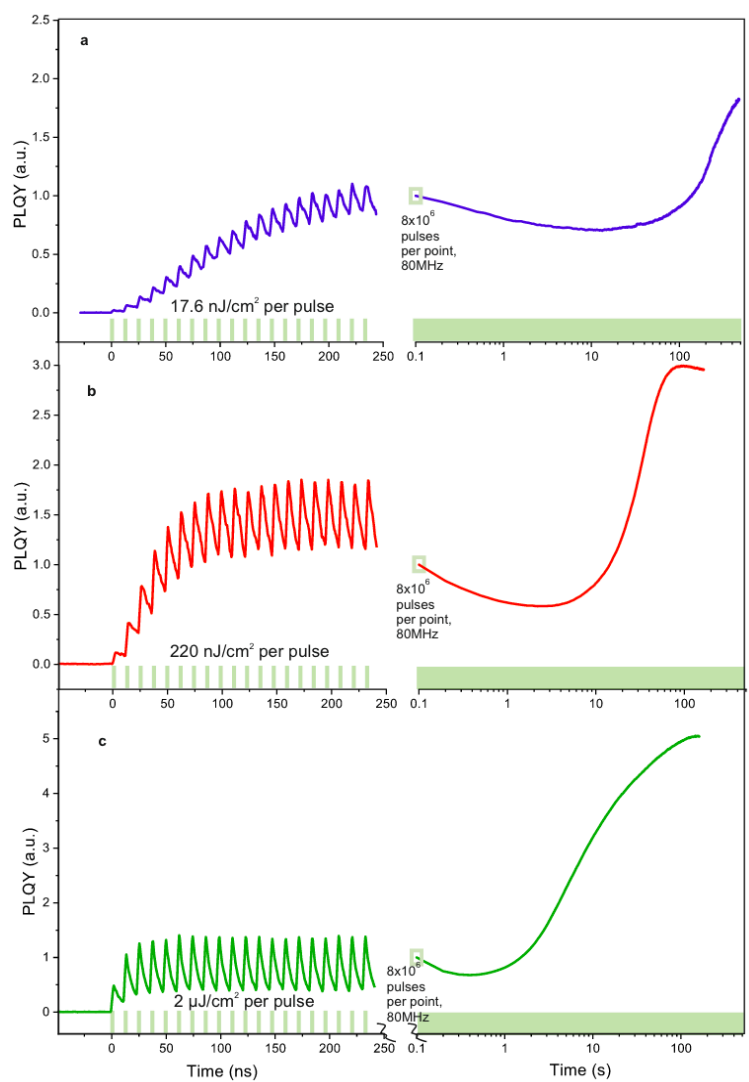

**Supplementary Fig. 9.** Evolution of the relative PLQY for different excitation conditions. a) Evolution of CsPbBr$_3$ film starting from $\vec{S}_1$ state under 17.6 nJ /cm$^2$ pulse fluence at 80 MHz pulse repetition. $\vec{S}_1$ state was prepared by exposing the sample to 2 µJ/cm$^2$@80 MHz (160 W/cm$^2$) irradiation for 20 seconds. b) and c) - Evolution of the sample starting from $\vec{S}_0$ state for at 220 nJ /cm$^2$ fluence and 2 µJ /cm$^2$ respectively.

It can be clearly seen that by increasing the pulse fluence, the shape of the leak-and-integrate behavior of PLQY changes. Initially, at the 17.6 nJ /cm$^2$ pulse fluence the on-off ratio between the 20$^{th}$ and the 1$^{st}$ pulse is high (**Supplementary Fig.9a)**, while for the 2000 nJ /cm$^2$, the saturation of the PL is quickly established after the 5$^{th}$ pulse (**Supplementary Fig.9c)**. In addition to this, a clear difference in the dynamics at the long timescale for the different pulse fluences is observed.

## Supplementary Note 6. Paired Pulse Facilitation of CsPbBr$_3$ and MAPbI$_3$ Memlumors

Here the synaptic plasticity for perovskite memlumors is demonstrated. By analogy to memristors, in which the synaptic weight is the conductance, in the memlumor case, the synaptic weight is the PLQY. The relative PLQY for each pulse can be obtained by dividing the PL output by the input light intensity. Thus, the difference between the PL of two nearby PL responses excited by the same input pulse intensity is the difference in the synaptic weight of a memlumor. By varying the delay time between the input pulses from the 12.5 ns to the 2.5 µs (**Supplementary Fig.10a)**, it is possible to demonstrate the typical paired pulse facilitation behavior for the CsPbBr$_3$ film (**Supplementary Fig.10b)** and the MAPbI$_3$ film (**Supplementary Fig.10c)** memlumors.

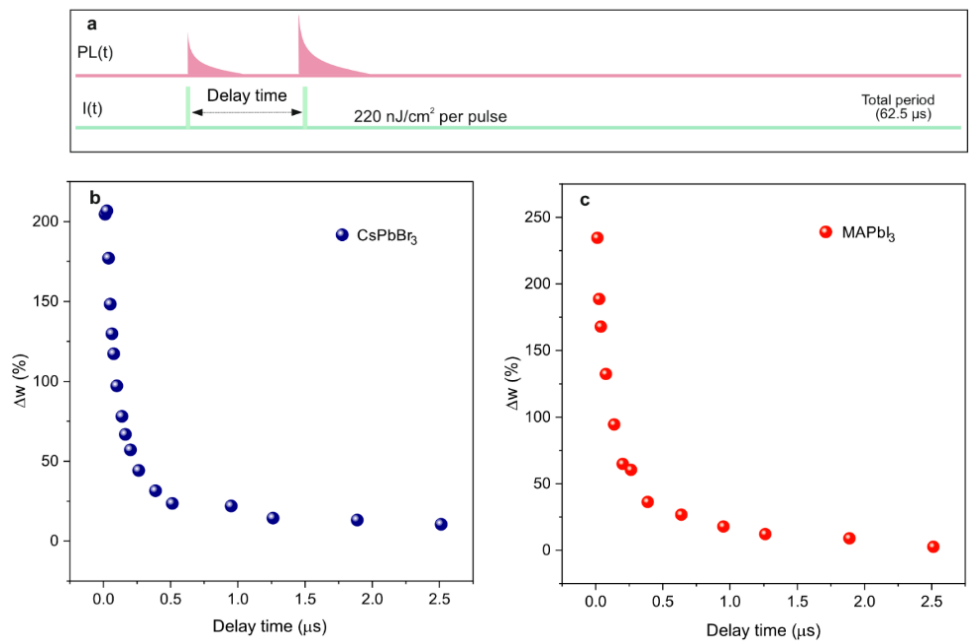

**Supplementary Fig. 10.** Measurement of the paired pulse facilitation (PPF). a) Typical experiment to detect PPF. When the delay time between the two nearby response pulses changes, the relative intensity of the responses also changes. PPF for CsPbBr$_3$ b) and MAPbI$_3$ c) film memlumors excited at the P3 fluence.

# Supplementary Note 7. Theoretical Modeling and Calculations

## S7.1. Shockley-Read-Hall (SRH) model with added radiative recombination (SRH+radiative)

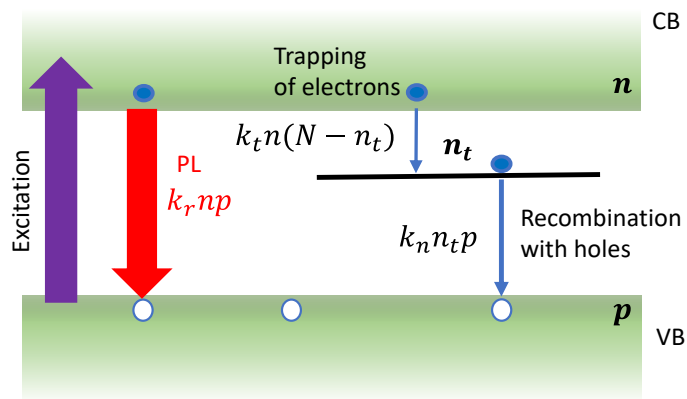

**Supplementary Fig. 11.** SRH charge recombination model with added radiative recombination (SRH+radiative). In this example the defect states are electron traps.

The SRH model considers carrier recombination occurring via trapping states located within the band gap [12]. An SRH model with added radiative term (SRH+radiative) can be used to explain the charge carrier recombination in a luminescent semiconductor [8]. The processes considered in the SRH model are illustrated in **Supplementary Fig.11** where an excess of charge carriers is generated by photoexcitation. These charge carriers can either recombine non-radiatively via the trap states (electron trapping followed by recombination with a free hole) or recombine radiatively leading to photoluminescence (PL) that can be detected experimentally. Mathematically, these processes can be described by a system of three differential equations. These equations describe the carrier density of electrons in the conduction band (n), holes density in the valence band (p) and density of trapped electrons ($n_t$) at the trap level:

$\frac{d}{dt}n(t) = G(t) - k_r np - k_t(N_t - n_t)n$     S7.1

$\frac{d}{dt}n_t(t) = k_t(N_t - n_t)n$     S7.2

$\frac{d}{dt}p(t) = G(t) - k_r np - k_n n_t p$     S7.3

where, G is the density of charge carriers generated by the excitation light per second, $k_r$ is the radiative recombination rate constant, $k_t$ is the electron trapping rate constant, $N_t$ is the electron trap density and, $k_n$ is non radiative recombination rate constant. The traps in the semiconductors are considered to be deep enough to make de-trapping of electrons negligible. Since an intrinsic semiconductor needs to be charge neutral, the condition of charge neutrality necessitates:

$n(t) + n_t(t) = p(t)$     S7.4

In presence of chemical doping the condition for charge neutrality becomes

$n(t) + n_t(t) + cd = p(t)$

here cd is the doping density, cd should be negative in case of n-doping or positive in case of p-doping.

The PL intensity in this model is provided as
$$PL(t) = k_r n(t) p(t) \qquad S7.5$$
Trapping of electrons by the trap states leads to an excess of holes in the valence band and results in the effect commonly referred to as photodoping.

In the equations above electron trapping was assumed. However, the same equations also apply for hole trapping by replacing n by p and *vise versa*. As a consequence, neither the model nor this type of experiments make it possible to distinguish between photodoping induced by electron trapping or by hole trapping.

## 7.2. Extended Shockley-Read-Hall model (SRH+) with added Auger processes.

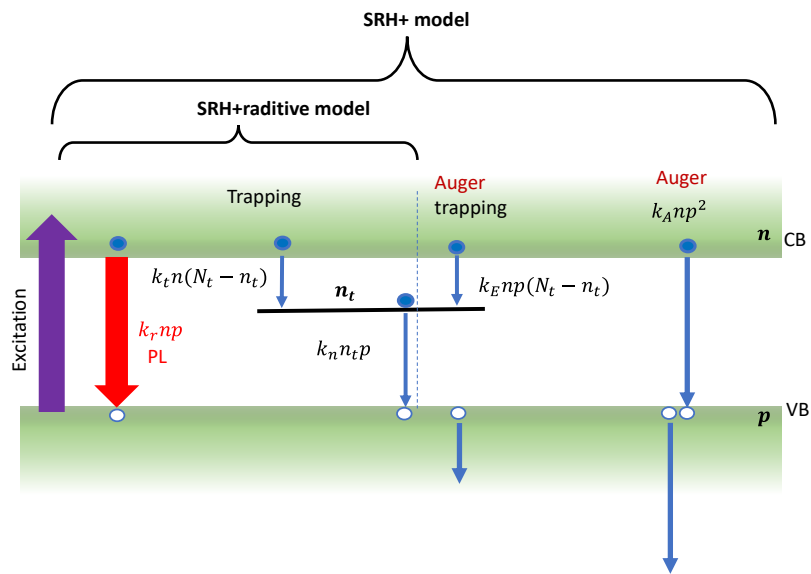

**Figure S12.** Schematic diagram of the processes considered in the SRH+ model, from left to right: photoexcitation, bimolecular radiative recombination, electron capture by the trap state, recombination of the trapped electron with a hole, Auger assisted electron capture (in the presence of an empty trap and a hole) and Auger assisted non-radiative recombination (electron recombines with a hole in the presence of another hole) [8]

Note that the model described above is only valid at low excitation conditions, such that the third order Augur assisted processes can be neglected. To explain the carrier recombination at high excitation conditions, the so-called SRH+ model[8] is applied. This model includes Auger recombination and Auger assisted charge trapping (**Supplementary Fig.12**).

The kinetic equations of the SRH+ model can be written as follows [8]:
$$\frac{d}{dt}n(t) = G(t) - k_r np - k_t(N_t - n_t)n - k_e np(N_t - n_t) - k_a np^2 \qquad S7.6$$
$$\frac{d}{dt}n_t(t) = k_t(N_t - n_t)n + k_e np(N_t - n_t) - k_n n_t p \qquad S7.7$$
$$\frac{d}{dt}p(t) = G(t) - k_r np - k_n n_t p - k_a np^2 \qquad S7.8$$
along with the condition of charge neutrality,
$$n(t) + n_t(t) = p(t)$$

Here, $k_e$, and $k_a$ are the Auger assisted electron capture rate constant and, Auger assisted recombination coefficient, respectively.

In this work, the SRH+ model was used to fit the PLQY (f,P) data and the PL decays (**Supplementary Fig.15**). These data sets were obtained for a very broad range of excitation conditions making it necessary to consider third order processes. Modeling of these experimental data made it possible to obtain all the parameters of the model (all rates and trap concentration) (**Supplementary Table 6**).

## 7.3. Pulsed photoexcitation in the framework of the SRH+ model

In a typical experiment, the PL is excited by a very short laser pulse at time t=0. Assuming that the pulse width is much shorter than all relaxation times in the system, the generation G(t) can be approximated as a delta function centered around t=0 creating the initial concentration of electron-hole pairs $n_0$. However, it is necessary to consider that practically no experiments are carried out with just one excitation pulse, instead the excitation pulse is repeated with a certain repetition period T. So, it is important to consider that the excitations created by the previous pulse may not have decayed at the moment of the next pulse's arrival. Under such circumstances, the charge carrier kinetic equations can be expressed as follows (see more details in SI to ref [8]:

$$\frac{d}{dt}n(t) = -k_r np - k_t(N_t - n_t)n - k_e np(N_t - n_t) - k_a np^2 \qquad \text{S.9}$$

$$\frac{d}{dt}n_t(t) = k_t(N_t - n_t)n + k_e np(N_t - n_t) - k_n n_t p \qquad \text{S7.10}$$

$$\frac{d}{dt}p(t) = -k_r np - k_n n_t p - k_a np^2 \qquad \text{S7.11}$$

with the periodic boundary conditions:
$$n(0) = n(T) + n_0$$
$$p(0) = p(T) + n_0$$
$$n_t(0) = n_t(T)$$

and the condition of charge neutrality
$$n(t) + n_t(t) = p(t)$$

Equations S7.9 to S7.11 can be numerically solved using a given periodic boundary to obtain the concentration of electrons and holes as a function of time. Subsequently, the value of PL quantum yield (PLQY which is measured experimentally) in a periodic pulsed excitation experiment can be calculated as:

$$PLQY = \frac{Number\ of\ photons\ emitted\ per\ one\ pulse}{number\ of\ e-h\ pairs\ created\ by\ one\ pulse} = \frac{k_r}{n_0}\int_{t'}^{t'+T} n(t)p(t)dt \qquad \text{S7.12}$$

where $t'$ is the starting time of the integration over the pulse repetition period. However, this starting time $t'$ must be chosen with care as is discussed below.

The distance between excitation pulses (excitation period T=1/f, where f - pulse repetition rate) in the experiments can be as short as 12.5 ns (80MHz repetition rate) or as long as hundreds of microseconds. If the repetition period of the excitation is so long that the densities of excess electrons, trapped electrons and excess holes decay to values close to zero, then this regime of excitation is called the *single pulse excitation regime*. It means that upon either a decrease or an increase in the excitation period, the PL response per one pulse should not change.

When the period becomes shorter, some excited species remain in the system when the next pulse arrives. Therefore, the PL dynamics starts to be dependent on the pulse period T. In the limiting case of a short distance between pulses, PLQY of the semiconductor starts to be dependent only on the time-averaged power density (W/cm$^2$), which is proportional to the product f*P. This regime is called quasi-CW excitation regime [8].

When the SRH+ model is applied in practice for calculating, *e.g.* PL(t) dynamics measured by TCSPC method, it is important to remember that experimentally this dependence is measured in the steady state condition in the sense that the sample is repeatedly excited with a laser pulse. Because the accumulation time of the PL decay (tens of seconds) is much larger than the time required for PL to reach quasi equilibrium, the TCSPC measures an equilibrated response of the system to a pulsed excitation.

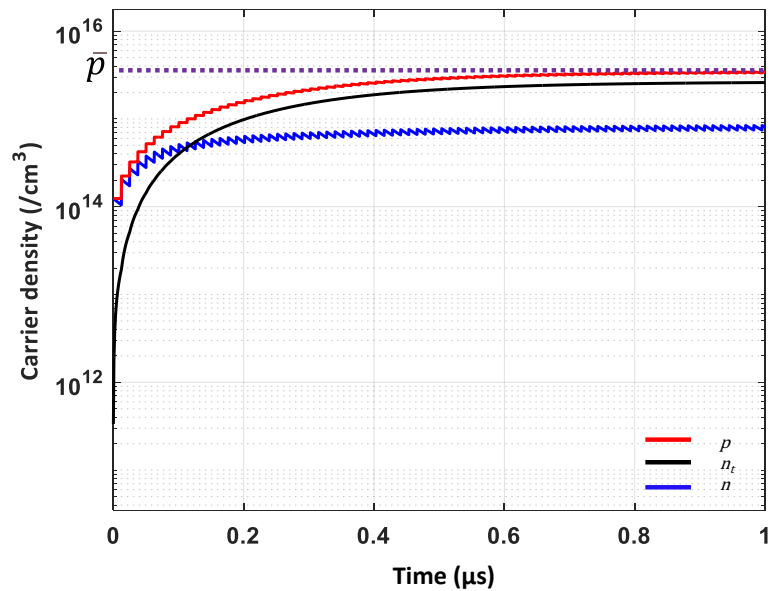

**Supplementary Fig. 13** Evolution of the electron density in conduction band (n), the hole density in valence band (p), and the trapped electron density ($n_t$) under continuous pulsed excitation with repetition rate 80 MHz and excitation power P2 (see **Supplementary Table 1**) which starts at time zero. Note the increase and the eventual saturation of $p \approx n_t = \bar{p}$. In this example the equilibrated concentration of free electrons is about 3 times less than of the free holes due to the photodoping effect. The time needed for saturation is about 1 μs.

If the excitation is in the single pulse regime, this equilibrated response is the same as the response to one single pulse. In other cases, one needs to apply multiple pulses until the system comes to the equilibrium. Experimentally, this is always the case because the accumulation time for PL decays is usually tens of seconds which is several orders of magnitude longer than the time required to reach equilibrium **(Supplementary Fig.13)**. To reproduce the experimental conditions in the calculations, numerically integrating the differential equations of the SRH+ model, makes it necessary to sequentially calculate as many repeated pulses as needed to establish the quasi-steady state values of the parameters.

This is illustrated in **Supplementary Fig.13** where one can see that the time-averaged concentrations of electrons (*n*) trapped electrons (*n$_t$*) and free holes (p) are increasing constantly after each excitation pulse until they reach certain equilibrated values. Note that between the pulses the concentrations still decrease, however, their dependences are stable and do not change anymore with applying any additional excitation pulses. This approach is utilized to calculate the PLQY value and PL decays in order to fit the PLQY(f,P) map (see **Supplementary Note 7.5 and 7.6**) and PL lifetime data.

Taking this into consideration, we have to conclude that the time $t'$ in equation S7.12 (the time from which the integration commences) should be larger than the time required for of the system to reach quasi-steady state condition. For calculations of the PL response to bursts of excitation pulses, the bursts were repeated several times with the needed repetition period until a quasi-steady state response is reached.

## 7.4 Solution of the model equations of SRH+ radiative model in the special case of low excitation fluence

An analytical solution of the equations of the SRH+ radiative model is possible at the limiting case of very low excitation density. In this case, the radiative recombination rate becomes negligible as compared to charge trapping rate. In addition, it is assumed that $n_t$ is significantly smaller than the trap density $N_t$ ($k_r p \ll k_t N_t$ and $n_t \ll N_t$). In this case, equations S7.1 – S7.3 can be approximated as:

$$\frac{d}{dt}n(t) = -k_t N_t n \qquad \text{S7.13}$$

$$\frac{d}{dt}n_t(t) = k_t N_t n - k_n n_t p \qquad \text{S7.14}$$

$$\frac{d}{dt}p(t) = -k_n n_t p \qquad \text{S7.15}$$

In this case, the periodic solution of S8.13 can be given as:

$$n(t) = \frac{n_0 \exp(-k_t N_t t)}{1-\exp(-k_t N_t T)} \qquad \text{S7.16}$$

**Single pulse regime under low excitation condition:**

If the period T is sufficiently large to satisfy
$$k_t N_t T \gg 1 \text{ and } k_n n_0 T \gg 1$$

the system is in the single pulse excitation regime.
From equation S8.16 it follows that:
$$n(t) = n_0 \exp(-k_t N_t t)$$

This indicates that the electron density in the conduction band drops exponentially with the characteristic time $1/k_t N_t$. It can be seen from equation S7.13 that the concentration of holes remains almost constant over the time it takes n(t) to fall by a factor of 1/e, provided that
$$t_d k_n n_0 = \frac{k_n n_0}{k_t N_t} \ll 1$$
In this case, one can substitute
$$p(t) = n_0$$
Thus, from equation S7.5 it follows that:
$$PL(t) = k_r n_0^2 \exp(-k_t N_t t) \qquad \text{S7.17}$$

and,
$$PLQY = \frac{k_r}{k_t N_t} n_0 \qquad \text{S7.18}$$

## 7.5 Modelling protocol for the extraction of the SRH+ model parameters

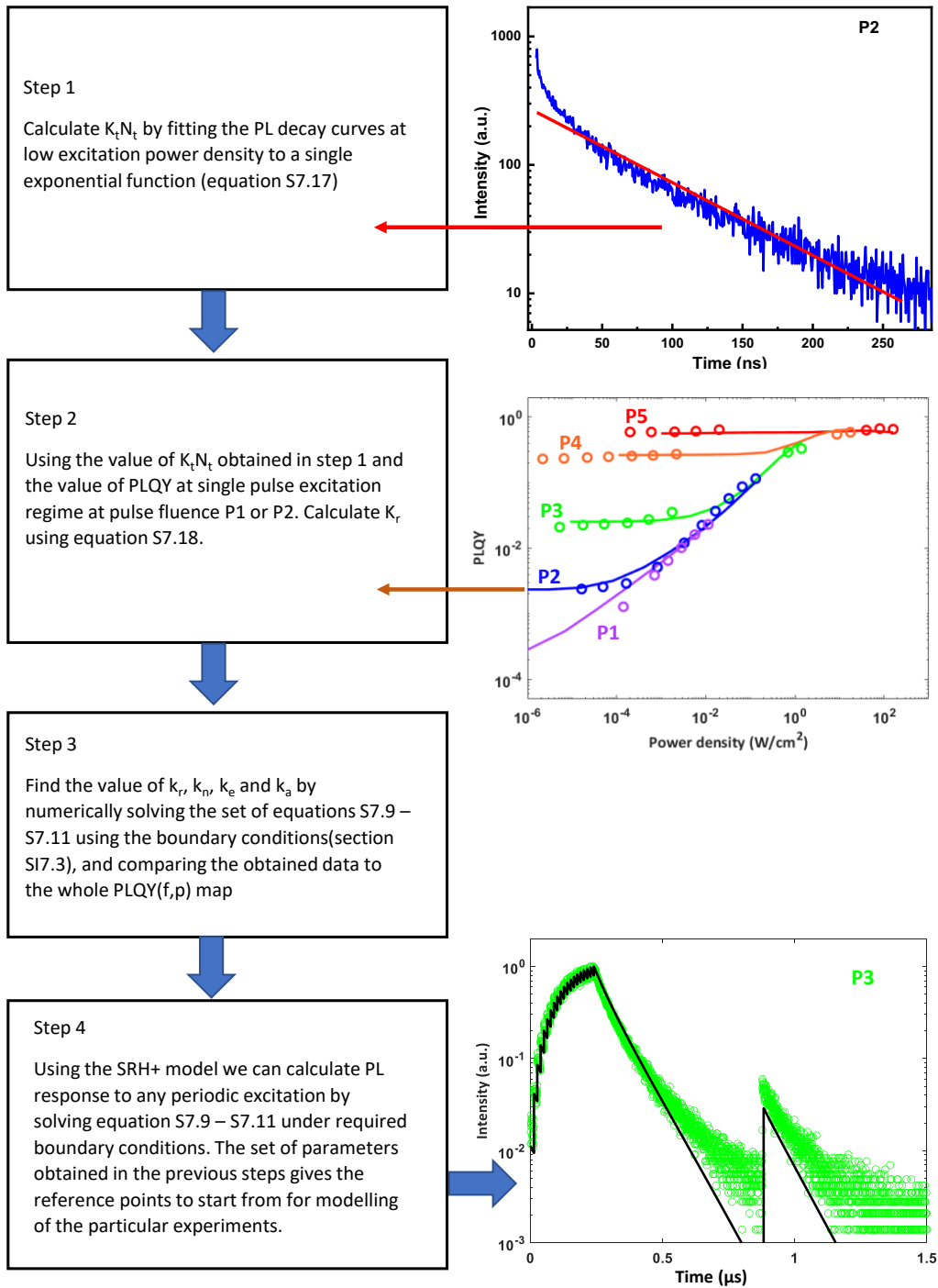

**Step 1**

Calculate $K_tN_t$ by fitting the PL decay curves at low excitation power density to a single exponential function (equation S7.17)

**Step 2**

Using the value of $K_tN_t$ obtained in step 1 and the value of PLQY at single pulse excitation regime at pulse fluence P1 or P2. Calculate $K_r$ using equation S7.18.

**Step 3**

Find the value of $k_r$, $k_n$, $k_e$ and $k_a$ by numerically solving the set of equations S7.9 – S7.11 using the boundary conditions(section SI7.3), and comparing the obtained data to the whole PLQY(f,p) map

**Step 4**

Using the SRH+ model we can calculate PL response to any periodic excitation by solving equation S7.9 – S7.11 under required boundary conditions. The set of parameters obtained in the previous steps gives the reference points to start from for modelling of the particular experiments.

**Supplementary Fig. 14.** Block diagram of the fitting procedure. The step 4 is illustrated by an example of a modelled experimental PL response to a write/read cycle, where the write is a burst of 20 pulses at P3 fluence and the read is a single pulse at the same fluence with a delay time of $\approx 0.6\ \mu s$ after the write (green – experiment, black – modelling). The parameters for the modelling were those obtained in Step 3 (**Supplementary Table 6**).

## 7.6. Extracting parameters according to the modeling protocol. Fitting experimental data for the CsPbBr$_3$ film memlumors.

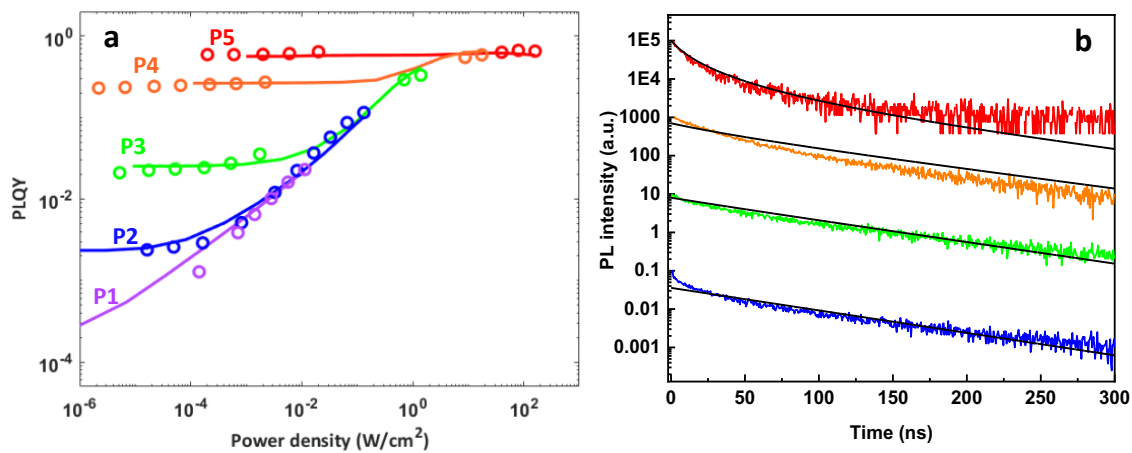

**Supplementary Fig. 15.** PLQY(f,P) map and the time-resolved PL for the CsPbBr$_3$ film and theoretical modelling of these data by the SHR+ model. a) The full PLQY(f,P) map. b) Time-resolved PL for different pulse fluences P1-P5 measured at 16 KHz repetition rate. The measurement of the PLQY(f,P) map is fully described in **Supplementary Note 2.5 and 4**.

**Supplementary Table 6.** The parameters of the SRH+ model for the CsPbBr$_3$ film memlumor (see **Supplementary Note 4**) obtained from the PLQY (f,P) mapping and PL decays (protocol **in Supplementary Note 7.5**). These parameters approximately correspond to the $\vec{S}_1$ state of the CsPbBr$_3$ film.

| | |
|---|---|
| $k_t N_t$ | $1.35 \times 10^7$ s$^{-1}$ |
| $N_t$ | $7.85 \times 10^{15}$ cm$^{-3}$ |
| $k_n$ | $7.93 \times 10^{-10}$ cm$^3$s$^{-1}$ |
| $k_r$ | $2.56 \times 10^{-10}$ cm$^3$s$^{-1}$ |
| $k_e$ | $2.10 \times 10^{-28}$ cm$^6$s$^{-1}$ |
| $k_a$ | $1.40 \times 10^{-27}$ cm$^6$s$^{-1}$ |

**Supplementary Table 7.** The parameters used to fit the experimental PL response of the $\vec{S}_1$ state of the CsPbBr$_3$ film memlumor in **Fig.3f** from main text.

| | |
|---|---|
| $k_t N_t$ | $1.35 \times 10^7$ s$^{-1}$ |
| $N_t$ | $7.85 \times 10^{15}$ cm$^{-3}$ |
| $k_n$ | $1.98 \times 10^{-9}$ cm$^3$s$^{-1}$ |
| $k_r$ | $2.56 \times 10^{-10}$ cm$^3$s$^{-1}$ |
| $k_e$ | $2.10 \times 10^{-28}$ cm$^6$s$^{-1}$ |
| $k_a$ | $1.40 \times 10^{-27}$ cm$^6$s$^{-1}$ |

**Supplementary Table 8.** The parameters used to fit the experimental PL response of the $\vec{S}_0$ state of the CsPbBr$_3$ film memlumor in **Fig.3f** from main text. For the modelling of the PL response in state $\vec{S}_0$ we changed only parameters k$_t$, k$_n$ and N$_t$ (components of the state vector $\vec{S}$) which are the properties of the traps, the other parameters here are the same as for the state $\vec{S}_1$ (**Supplementary Table 7**).

| k$_t$N$_t$ | 2.43x10$^7$ s$^{-1}$ |
|---|---|
| N$_t$ | 7.85x10$^{15}$ cm$^{-3}$ |
| k$_n$ | 1.19x10$^{-8}$ cm$^3$s$^{-1}$ |
| k$_r$ | 2.56x10$^{-10}$ cm$^3$s$^{-1}$ |
| k$_e$ | 2.10x10$^{-28}$ cm$^6$s$^{-1}$ |
| k$_a$ | 1.40x10$^{-27}$ cm$^6$s$^{-1}$ |

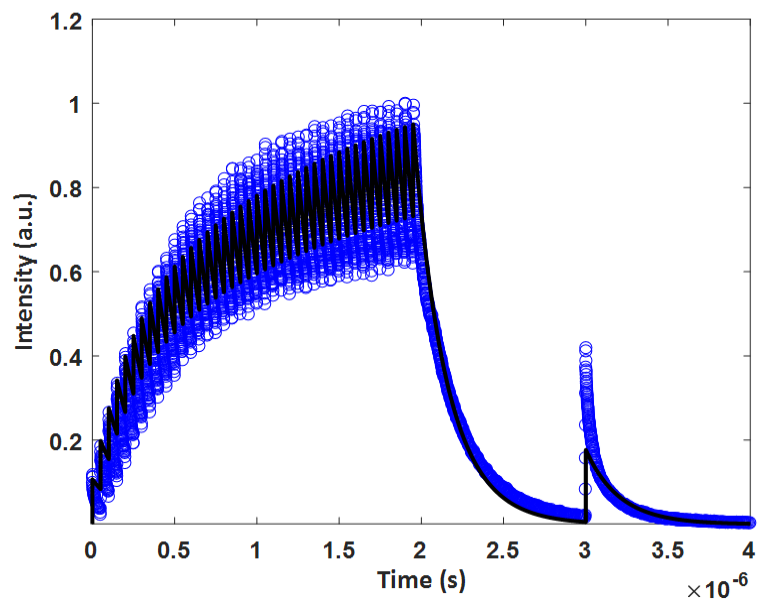

**Supplementary Fig. 16.** Modelling of the PL response shown in **Figure 2d** in the main text (writing with a burst of 40 pulses (P2 fluence) at 20 MHz and reading with 1 pulse). Blue circles – experiment, black line – modelling.

**Supplementary Table 9.** Parameters used to model the experimental PL response the CsPbBr$_3$ film memlumor shown in **Supplementary Fig.16** and **Fig.2d** in main text. Here k$_r$, k$_e$ and k$_a$ are the same as obtained from the PLQY(f,P) mapping (see **Supplementary Note 4, S7.5**) while the parameters of the vector $\vec{S}$ (k$_t$, N$_t$, k$_n$) are modified. Also, a permanent excess of holes (chemical doping, c$_d$) was added to make the model able to better mimic the experiment. This c$_d$ parameter accounts for the so-called accumulating effect discussed in the main text.

| k$_t$N$_t$ | 4.05x10$^6$ s$^{-1}$ |
|---|---|
| N$_t$ | 1.18x10$^{16}$ cm$^{-3}$ |
| k$_n$ | 7.23x10$^{-11}$ cm$^3$s$^{-1}$ |
| k$_r$ | 2.56x10$^{-10}$ cm$^3$s$^{-1}$ |
| k$_e$ | 2.1x10$^{28}$ cm$^6$s$^{-1}$ |
| k$_a$ | 1.40x10$^{-27}$ cm$^6$s$^{-1}$ |
| c$_d$ | 2.97x10$^{15}$ cm$^{-3}$ |

## 7.7 Modelling of short-term memory effect by modifying parameters of trap states

**Supplementary Fig.17, 18 and 19** demonstrate how the capacity of a system to hold memory can be tuned by changing the vector $\vec{S} = (k_t, k_n, N_t)$, and also by choosing an appropriate energy of the writing pulses. For these calculations, a writing cycle by 400 pulses with 80 MHz repetition rate (12.5 ns between pulses) was selected. The memory effect is characterized by the ratio:

$$\frac{W_{400}}{W_0} = \frac{\text{PL excited by single pulse after writing with 400 pulses}}{\text{PL excited by single pulse without writing}}$$

where the PL means the PL signal integrated over the entire PL decay generated by one pulse. The larger this ratio, the more information can be stored in the system.

The PL response shown in **Supplementary Fig.17, Supplementary Note 18 and 19** is calculated for the pulse energies $P_1/100$, $P_1$ and $P_3$, respectively with the pulse repetition rate set to 80 MHz. The panel (a) in each figure illustrates the PL response with the parameters extracted from the PLQY(f,P) and PL decay measurements (see **Supplementary Note 7.5, 7.6 and Supplementary Table 6**), these parameters are labeled with index zero. The other three panels show the effect of decreasing by a factor of 5 of various parameters, i.e.: b) trap concentration $N_t$, c) electron trapping rate $k_r$ and d) recombination rate $k_n$ between a trapped electron and a free hole.

As one can see the largest effect is observed by changing the recombination rate $k_n$ between the trapped electrons and free holes and the pulse fluence. For the lowest $k_n$ and the lowest pulse fluence it was possible to obtain $W_{400}/W_0$ as large as 381.

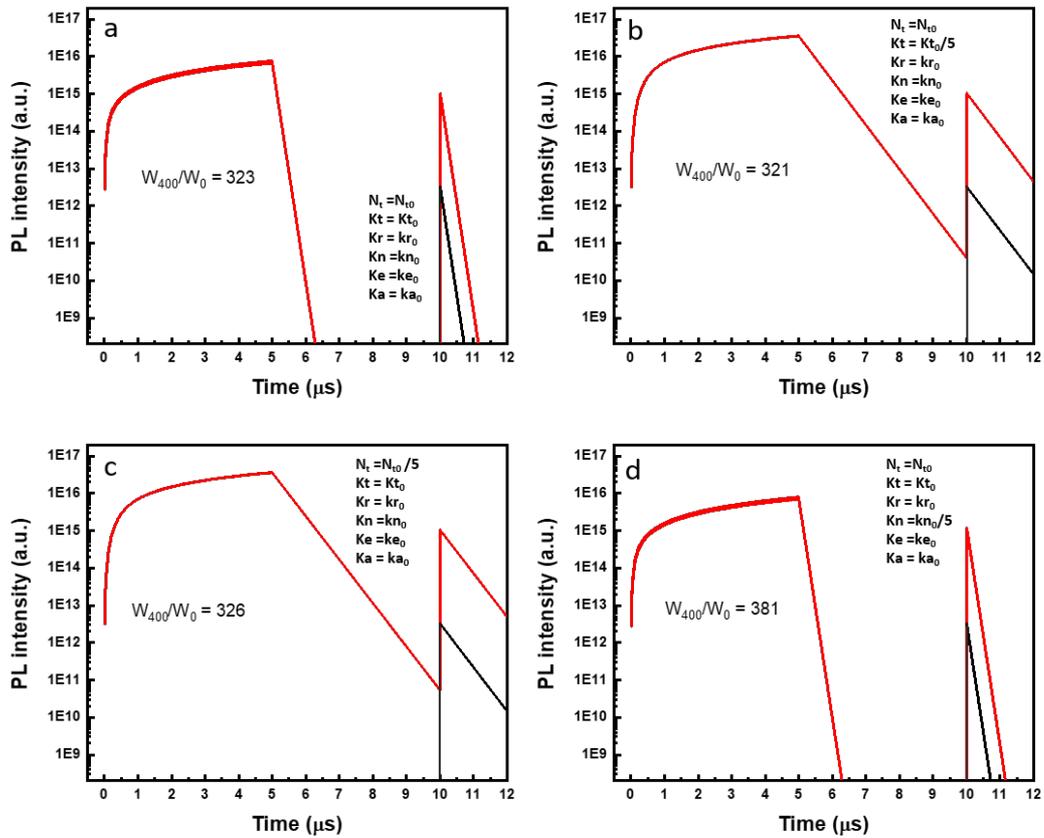

**Supplementary Fig. 17.** Simulated PL response of CsPbBr$_3$ on the writing/reading cycle under pulse fluence P$_1$/100, the ratio W$_{400}$/W$_0$ is indicated. a) standard model parameters from **Supplementary Table 6**. b) k$_t$ is decreased by a factor of 5, c) N$_t$ decreased by a factor of 5, d) k$_n$ decreased by a factor of 5.

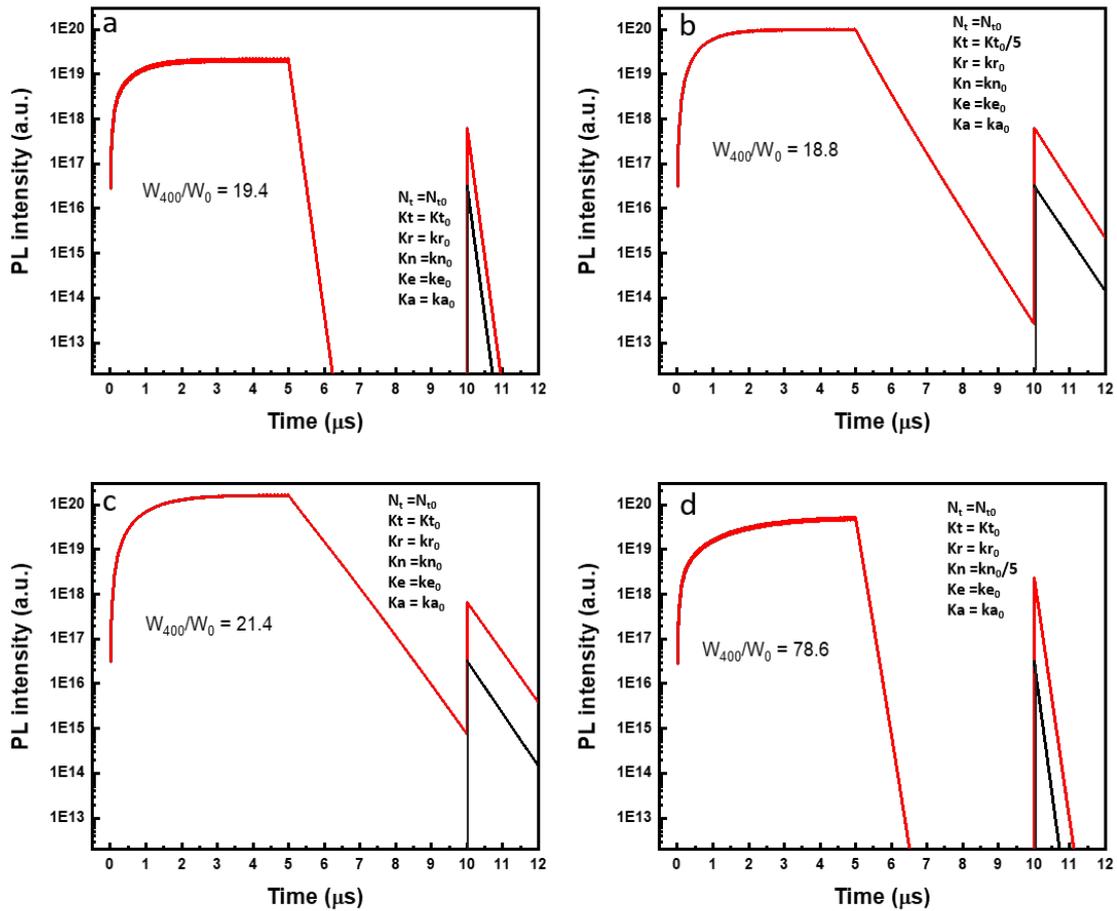

**Supplementary Fig. 18.** Simulated PL response of CsPbBr$_3$ on writing/reading for the pulse fluence P$_1$, the ratio W$_{400}$/W$_0$ is indicated. a) and standard model parameters (see **Supplementary Table 6**): b) k$_t$ is decreased by a factor of 5, c) N$_t$ decreased by a factor of 5, d) k$_n$ decreased by a factor of 5.

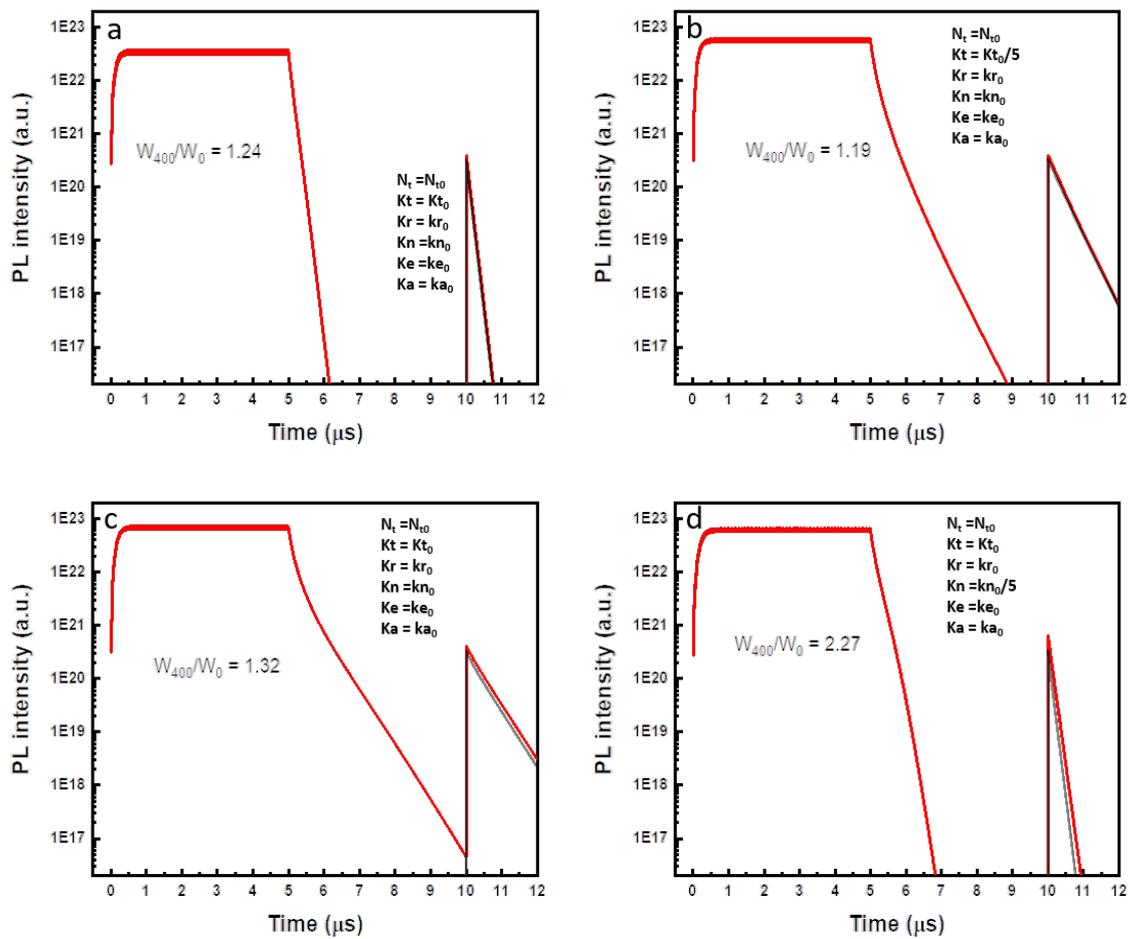

**Supplementary Fig. 19.** Simulated PL response of CsPbBr$_3$ on writing/reading for the pulse fluence P3 the ratio W$_{400}$/W$_0$ is indicated. a) and standard model parameters (see **Supplementary Table 6**): b) k$_t$ is decreased by a factor of 5, c) N$_t$ decreased by a factor of 5, d) k$_n$ decreased by a factor of 5.

# Supplementary Note 8. CsPbBr$_3$ Memlumor Crystals Integrated on GaP Waveguides

To demonstrate the summation operation by memlumors, sub-micrometer sized CsPbBr$_3$ crystals were integrated with GaP waveguides. For the signal collected from the edge of the waveguide, the synaptic weight is a property of the size of each crystal, its own PLQY and the waveguiding efficiency. The latter means that the same crystal located at different places of the waveguide should result in a different signal magnitude.

By diaphragming the excitation light input, it is possible to address independently different memlumors in the complex system (**Supplementary Fig. 20a**, **Supplementary Fig. 20b**) or illuminate them all at once by a broad excitation beam (**Supplementary Fig. 20c**). In the latter case, the signal collected at the edge is a summation of the PL from all crystals guided to the waveguide edge.

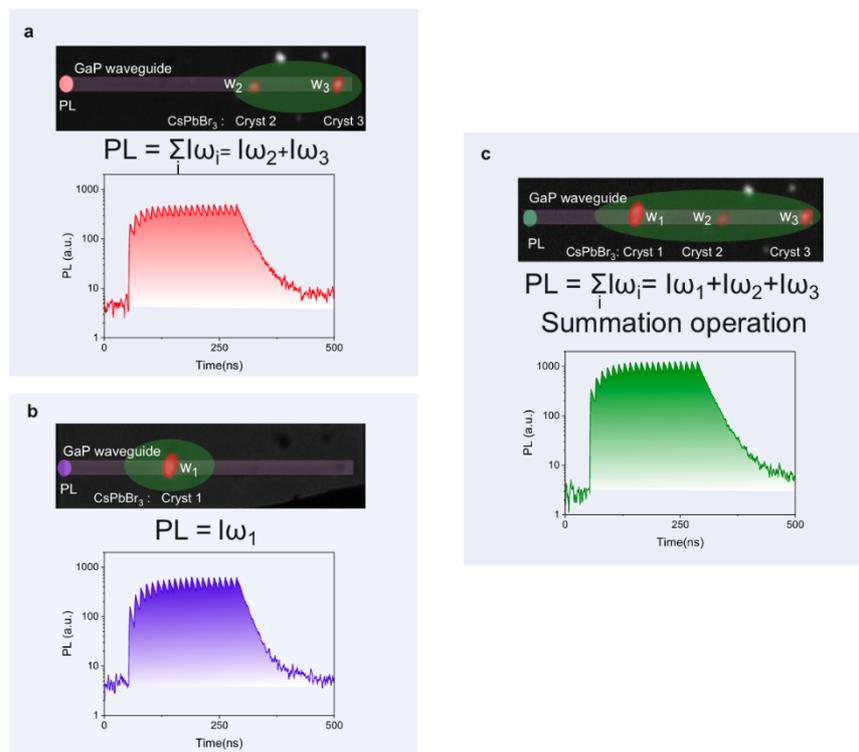

**Supplementary Fig. 20.** Experiment with a GaP waveguide. Three CsPbBr$_3$ crystals are attached to the GaP waveguide. Their PL propagates in the waveguide to be collected from its edge. The light spot (green) can be controlled so it can illuminate 1, 2 or 3 crystals (panels a,b and c, respectively).